\documentclass[
aps,prd,
showpacs,reprint,notitlepage,
amssymb,amsmath,amsfonts,mathrsfs,
nofootinbib,superscriptaddress,
floats,floatfix
]{revtex4-2}

\usepackage{graphicx}
\usepackage{xcolor}
\usepackage[colorlinks=true]{hyperref}
\hypersetup{citecolor=cyan,linkcolor=magenta}
\usepackage{multirow,array}
\DeclareMathAlphabet{\pazocal}{OMS}{zplm}{m}{n}
\usepackage{journals}

\newcommand{\kom}{M_{\rm K}}
\newcommand{\Oo}{\Omega_{\rm orb}}
\newcommand{\comp}{\lambdabar_{\rm comp}}
\newcommand{\Ms}{M_{\star}}
\newcommand{\Rs}{R_{\star}}
 
\definecolor{maroon}{cmyk}{0,0.87,0.68,0.32}
 
\newcommand{\nn}{\nonumber\\}

\begin{document}
\title{Binary neutron star mergers in massive scalar-tensor theory: Quasi-equilibrium states and dynamical enhancement of the scalarization}

\date{\today}

\author{Hao-Jui Kuan}
\email{hao-jui.kuan@aei.mpg.de}
\affiliation{Max Planck Institute for Gravitational Physics (Albert Einstein Institute), 14476 Potsdam, Germany}

\author{Karim Van Aelst}
\affiliation{Max Planck Institute for Gravitational Physics (Albert Einstein Institute), 14476 Potsdam, Germany}

\author{Alan Tsz-Lok Lam}
\affiliation{Max Planck Institute for Gravitational Physics (Albert Einstein Institute), 14476 Potsdam, Germany}

\author{Masaru Shibata}
\affiliation{Max Planck Institute for Gravitational Physics (Albert Einstein Institute), 14476 Potsdam, Germany}
\affiliation{Center of Gravitational Physics and Quantum Information, Yukawa Institute for Theoretical Physics, Kyoto University, Kyoto, 606-8502, Japan} 

\begin{abstract}
We study quasi-equilibrium sequences of binary neutron stars in the framework of Damour-Esposito-Farese-type scalar-tensor theory of gravity with a massive scalar field, paying particular attention to the case where neutron stars are already spontaneously scalarized at  distant orbits, i.e., in the high coupling constant case. Although scalar effects are largely quenched when the separation $a$ is $\agt 3$--$6$ times of the Compton length-scale that is defined by the scalar mass, we show that the interaction between the scalar fields of the two neutron stars generates a scalar cloud surrounding the binary at the price of orbital energy when $a \alt 3$--$6$ times of the Compton length-scale. This enables us to constrain the scalar mass $m_\phi$ from gravitational-wave observations of binary neutron star mergers by inspecting the dephasing due to such phenomenon. In particular, the event GW170817 is suggestive of a constraint of $m_\phi \agt 10^{-11}$~eV and the coupling strength should be mild if the neutron stars in this system were spontaneously scalarized. 
\end{abstract}

\maketitle

\section{Introduction}\label{secI}

General relativity (GR) has been put against a variety of observations and yet been challenged, while it has also proven to be incomplete from the theoretical point of view for its nonrenormalizability (e.g., \cite{bart83,barv85}). Among the extensions to GR present in the literature, Damour-Esposito-Farese (DEF) type of scalar-tensor (ST) theory of gravity is perhaps most widely considered. In such theory, the gravity around a scalarized compact object acquires distinct feature from that in GR, modifying the trajectory of orbiting companions.
In particular, the motion of binaries will be influenced to deviate from the GR prediction if there is scalar interaction between the two components at play. In addition, scalar waves will be emitted from binaries consisting of differently scalarized components, constituting extra loss of orbital energy.
Lacking the evidences of the aforementioned two effects in the pulsar timing observation of neutron star-white dwarf (NS-WD) binaries has placed strong constraints on ST theories with a {\em massless} scalar field \cite{damo98,frei12,shao17,ande19,guo21,beni23}. Such constraints are rather stringent for the presence of a scalar charge of neutron stars (NSs)~\cite{chib22,zhao22}. These constraints can, however, be mitigated by the inclusion of scalar mass $m_\phi$ \cite{alsi12,rama16}. The scalar effects beyond the associated Compton length-scale $\comp=\hbar c/m_\phi$ are smeared out, thus naturally accounting for the non-detection of scalar dynamics that could take place in these binaries. In particular, the constraints by the pulsar timing are lifted to a large extent if the scalar field has a light mass $m_\phi\gg 10^{-16}$~eV (corresponding to a Compton length-scale $\comp\ll1.5\times10^{6}$~km) \cite{rama16}. With this small mass, the scalar interaction within NS-WD binaries and the emission of scalar waves from them are suppressed, leading to the identical orbital evolution with that in GR. Therefore, including a scalar mass not only increases the dimension of the parameter space by one, but unlocks the previously ruled out region. 
However, NS-WD binaries could barely put constraints on the massive theory since a light scalar field is enough to lift the constraining power of pulsar timing observations. On the other hand, an ever-stringent lower bound on the scalar mass may be placed by pre-merger gravitational waves (GWs) from coalescing binary neutron stars (BNSs).

For BNS mergers, the growth of the scalar field can be activated by the gravitational compactness of the binary, defined as the ratio of the total mass to the orbital separation, forming another kind of scalarization \cite{bara13,shib13,shib14,samp14} (see also \cite{pale14,senn16,khal19,khal22} for semi-analytical modeling) other than the spontaneous ones  \cite{damo93,andr19}. In the same spirit as pulsar timing constraints, the absence of both kinds of scalarization in the event GW170817 suggests that spontaneously scalarized NSs are unlikely present in the associated coalescing BNS if the scalar field is \emph{massless} \cite{zhao19}. To probe massive ST theory by GW physics, a pursue of scalar masses $10^{-12}$--$10^{-11}$~eV is of particular interest since the associated Compton length-scale is comparable with or smaller than the typical orbital separation of $\sim 30$--$200$\,km when the BNS comes in the detection window.

It is widely known that the uncertainty on the theory of gravity is degenerate with that on the nuclear equations of state (EOS) \cite{sota17,shao19}. Among other things, the twin star in GR predicted from some EOS embracing hadron-quark phase transition has an analog in the ST theory \cite{kuan22}. Nonetheless, certain scalar-induced phenomena have no counterparts in GR, e.g., the presence of scalar-type GWs from binary motions \cite{damo92}, core-collapse of giant stars \cite{sper17,cheo19,geng20b,rosc20,asak23}, and radial \cite{sota14} and polar \cite{krug21} oscillations of NSs (see \cite{done23} for a recent, extensive review). An observation of such ST-exclusive effects can therefore probe the nature of gravity, and limit the parameter space of ST theories without the potential for misinterpreting EOS effects. The dynamics during the late inspiral up to merger, and the associated GW emission from BNSs in a ST theory that admits spontaneous and/or dynamical scalarization may shed unique light on the nature of gravity \cite{abbo17,abbo19}, thus deserving qualitative investigation.

For mass of $m_\phi \agt 10^{-12}$~eV, the scalar effects are shielded in the early inspiral and the interaction only becomes dynamically important when the binary approaches merger. 
Since the effects occur in a highly non-linear regime of the theory, it can only be investigated numerically. Although certain attempts have been made in the massless case ($m_\phi=0$) \cite{heal12,bara13,shib14,tani14,ma23}, numerical study of the BNS dynamics in theories with a massive scalar field has not been performed. We thus endeavour to address such issue numerically as a non-trivial scalar mass is necessary to account for the aforementioned observations. For this purpose, preparing appropriate initial data (ID) is rather imperative in order to guarantee accurate simulations.

As the first step towards the derivation of accurate BNS dynamics and the emitted GWs, we develop an ID code to generate equilibrium states of BNSs, which are expected to deliver certain information on the dynamics of coalescence since the sequence of equilibria can be viewed as the leading order approximations of the inspiraling process. In particular, the constructed equilibria can (i) offer an approximate estimate on the luminosity of GWs \cite{shib01,tani14}, and (ii) qualitatively investigate scalar effects in the inspiral stage on top of (iii) paving the way toward future numerical-relativity studies of BNS mergers. By scrutinising the constructed sequences, we found that a lower bound of $m_\phi>10^{-11}$~eV for strong couplings can be readily drawn. Although quantitative analysis of the waveforms can supplement the effort of waveform-modelling (e.g., \cite{boni23}) to examine the imprint of modified gravity from GWs, the relevant investigation will be deferred to later work in this series.

In this paper, we pay particular attention to the sequences of BNSs in which each NS is spontaneously scalarized, i.e., the coupling constant $B$ is high [see Eq.~(\ref{eq2})]. Broadly speaking, inspiraling scalarized BNSs are speculated to be classified into three stages depending on the following three parameters: the orbital separation $a$, the gravitational wavelength $\lambdabar_\mathrm{gw}$, which is $\approx a^{3/2}M^{-1/2}/2 (>a)$ for binaries in circular orbits with $M$ the total mass of the binary, and the Compton length-scale $\comp$. 
For (I) $\lambdabar_\mathrm{gw} > a \gg \comp$, no effect associated with the scalar field appears and hence the sequences of BNSs can be identical to those in GR; (II) for $\lambdabar_\mathrm{gw} > \comp \agt a$, the scalar-wave emission is suppressed because of the relation $\lambdabar_\mathrm{gw} > \comp$, while the interaction between the scalar clouds of the two NSs can play a role in modifying the binary orbit; (III) for $\comp> \lambdabar_\mathrm{gw} > a$, both the scalar-wave emission and interaction of the two scalar clouds are present. For the categories (II) and (III), the orbital evolution of the BNSs can be different from that in GR. One of the primary purposes of this paper is to confirm these speculations.

This paper is organized as follows. Section \ref{secII} briefly reviews the ST theory studied, including the connection to other formalisms adopted in the literature, the definition of 'mass', and constraints on the theory parameters from current observations of binary pulsar timing and GWs from coalescing BNS. In Sec.~\ref{secIII}, we construct sequences of quasi-equilibrium states and elaborate on the novel phenomenon coined as scalar-induced plunge. Discussion and potential implications of a detection of such effects are given in Sec.~\ref{final}.
Throughout this paper, we adopt the geometric units, i.e., $G=1=c$, together with the reduced Plank constant set to $\hbar=1$. The indices $a$, $b$, and $c$ denote the spacetime components and $i$, $j$, and $k$ the spatial components.

\section{Theoretical and observational aspects of the theory}\label{secII}

\subsection{Basic equations}

The action of the scalar-tensor theory in the Jordan frame is written as \cite{jord59,bran61}
\begin{align}
    S =& \frac{1}{16\pi}\int d^4x \sqrt{-g}
    \left[ \phi {\cal R} - \frac{\omega(\phi)}{\phi}
    \nabla_a \phi \nabla^a \phi - U(\phi) \right] \nonumber\\
    &- \int d^4x \sqrt{-g} \rho (1 + \varepsilon),
\end{align}
where ${\cal R}$ and $\nabla_a$ are the Ricci scalar and covariant derivative associated with the metric $g_{ab}$, $\rho$ is the rest-mass density, and $\varepsilon$ is the specific internal energy. In the action, $\omega(\phi)$ describes the coupling between the metric and the scalar field $\phi$, for which the following expression:
\begin{align}
    {1 \over \omega(\phi)+3/2}= B\, \ln \phi, \label{eq2}
\end{align}
is adopted in the present article with $B$ as the dimensionless coupling constant~\cite{shib14}. For latter use, we introduce the variable $\varphi$ via
\begin{align}\label{eq:defvarphi}
    2\ln\phi = \varphi^2,
\end{align}
with respect to which the scalar potential,
\begin{align}\label{eq:potential}
    U(\phi)=\frac{2m_\phi^2\varphi^2\phi^2}{B},
\end{align}
is chosen for the scalar mass $m_\phi$ \cite{kuro23}. Along with the scalar mass, a Compton length-scale,
\begin{align}
    \comp \approx 19.7\,\mathrm{km} \left( \frac{m_\phi}{10^{-11}\,\,{\rm eV}} \right)^{-1}
\end{align}
is introduced.

Denoting the Einstein tensor associated with the metric $g_{ab}$ as $G_{ab}$, the equation of motion associated with the action can then be written down as
\begin{align}
    G_{ab} =& 8\pi \phi^{-1}T_{ab} + \omega(\phi)\phi^{-2}\left[ \nabla_a\phi\nabla_b\phi-\frac{1}{2}g_{ab}\nabla_c\phi\nabla^c\phi \right] \nn 
    &+\phi^{-1}(\nabla_a\nabla_b\phi-g_{ab}\nabla_c\nabla^c\phi) - \frac{2m_\phi^2}{B}\phi\ln\phi g_{ab},
\end{align}
and 
\begin{align}\label{eq:KG}
    \nabla_a \nabla^a \phi = \frac{1}{2\omega(\phi)+3}\left[ 8\pi T -\frac{d\omega}{d\phi}\nabla_c\phi\nabla^c\phi +\frac{4m_\phi^2\phi^2}{B} \right],
\end{align}
where $T_{ab}$ is the stress-energy tensor and $T=T_a^{~a}$. The equation of motion for the matter in the Jordan frame is the same as in GR, i.e.,
\begin{align}
    \nabla_aT^{ab}=0.
\end{align} 
The fluid is assumed to be a perfect fluid, for which the stress-energy tensor has the form
\begin{align}
    T^{ab}=\rho hu^au^b + Pg^{ab},
\end{align}
where $P$ is the pressure, $h=1+\varepsilon+P/\rho$ is the specific enthalpy, and $u^a$ is the 4-velocity of the fluid, respectively.

\subsection{Connection to the Einstein frame}
To draw the connection to a large part of the literature, where the Einstein frame is often considered due to certain advantages with respect to the Jordan frame, we provide the relations between these two frames in this subsection, while we will stick to the Jordan frame in the rest of the article. The scalar field in the Einstein frame, denoted by $\bar{\varphi}$, is defined by assuming that the Weyl relation between the metric fields in the two frames is
\begin{align}\label{eq:weyl}
    g_{ab} &= A(\bar{\varphi})^2 g_{ab}^E, 
\end{align}
where $A(\bar{\varphi})=\phi^{-1/2}=e^{\beta_0\bar{\varphi}^2/2}$, and $\beta_0$ is a dimensionless constant. Thus, 
\begin{align}\label{eq:j2e}
    \varphi=\sqrt{-2\beta_0}\bar{\varphi}=\sqrt{B}\bar{\varphi}.
\end{align}
In addition, the potential in the Einstein frame, $V$, related to $U$ via $U=4V\phi^2$, is given by
\begin{align}
    V=\frac{1}{2}m_\phi^2\bar{\varphi}^2,
\end{align}
which makes clear the physical meaning of the parameter $m_\phi$ as the scalar mass.

The two parameters in the DEF theory are defined as the asymptotic values of the first and second derivative of the logarithmic coupling function \cite{damo92,damo93}. Let the asymptotic value of the Jordan frame scalar field be $\varphi_0$, thus the one in the Einstein frame being $\bar{\varphi}_0=\varphi_0/\sqrt{B}$ by Eq.~\eqref{eq:j2e}, one then has 
\begin{align}
	\alpha_{\rm DEF}= \left.\frac{d\ln A}{d\bar{\varphi}} \right|_{\varphi_0} =\frac{\beta_0\varphi_0}{\sqrt{B}},
\end{align}
and
\begin{align}
	\beta_{\rm DEF}=\left.\frac{d^2\ln A}{d\bar{\varphi}^2} \right|_{\varphi_0} =\beta_0=-B/2.
\end{align}
As long as the transformations of the fields between the two frames are mathematically well-defined (e.g., one-to-one relations should be guaranteed \cite{geng20}), the physics can be equally validly discussed in whichever frame \cite{flan04}.

\subsection{Gravitational field equations in quasi-equilibria}

We describe here the basic gravitational field equations for computing quasi-equilibria of BNSs in circular orbits. Following previous works~\cite{isen08, wils89} (and see,  e.g.,~\cite{tani10} for a review), we solve the constraint equations under the maximal slicing condition, assuming conformal flatness for the 3-spatial metric $\gamma_{ij}=W^{-2} f_{ij}$, where $W$ is a conformal factor and $f_{ij}$ is the flat 3-metric. 

The momentum constraint is written as
\begin{align}
    0={\cal M}_j&=D_iK^i{}_j-D_jK \nonumber 
    -8\pi\phi^{-1}J_j +\varphi K_j{}^i D_i\varphi \\
    & -\left(1+ \frac{2}{B}-\frac{\varphi^2}{2} \right)\Phi D_j\varphi
    -\varphi D_j\Phi,
\end{align}
where $D_i$ denotes the covariant derivative with respect to $\gamma_{ij}$, $K_{ij}$ is the extrinsic curvature with $K=K_k^{~k}$, $\Phi=-\alpha^{-1}(\partial_t -\beta^k \partial_k)\varphi$ 
with $\alpha$ the lapse function and $\beta^k$ the shift vector, and $J_i=\alpha T_i^{~t}$. 
The Hamiltonian constraint is written as
\begin{align}
    0={\cal H}&=R+K^2-K_{ij}K^{ij}-16\pi\phi^{-1}\rho_{\rm h} \nonumber\\
    &-\left( \frac{2}{B}-\frac{3}{2}\varphi^2 \right)(\Phi^2+D_k\varphi D^k\varphi) \nonumber\\
    &-2\left[-K\Phi\varphi+\varphi D_kD^k\varphi+(1+\varphi^2)D_k\varphi D^k\varphi\right] \nonumber\\
    &-\frac{2m_\phi^2\varphi^2\phi}{B},
\end{align}
where $R$ is the Ricci scalar with respect to $\gamma_{ij}$ and $\rho_\mathrm{h}=\alpha^2 T^{tt}$. 

The elliptic equations for generating binary ID (assuming conformal flatness) are written down as (see \cite{tani14,shib14} for equations in ST theories with a massless scalar field)
\begin{align}\label{eq:elp1}
    \Delta\psi=&-\phi^{-1}\psi^5\left( 2\pi\rho_{\rm h}+\frac{m_\phi^2\phi^2\varphi^2}{4B} \right)-\frac{1}{8}\psi^{-7}\bar{A}_{ij}\bar{A}^{ij} \nonumber\\
    &-\frac{1}{2}\pi B\psi^5\varphi^2 T \phi^{-1} - \frac{m_\phi^2\phi \varphi^2}{4}\psi^5 \nonumber\\
    &-\frac{\psi}{4}\left( 1+\frac{1}{B}-\frac{3}{4}\varphi^2 \right)f^{ij}(\partial_i\varphi)(\partial_j\varphi) \nonumber\\
    &+\frac{1}{4}\chi^{-1}\varphi f^{ij}(\psi\partial_i\chi-\chi\partial_i\psi)(\partial_j\varphi),
\end{align}
\begin{align}\label{eq:elp2}
    \Delta\chi
    =&\,\,2\pi\phi^{-1}\chi\psi^{4}(\rho_{\rm h}+2S)
    +\frac{7}{8}\chi\psi^{-8}\bar{A}_{ij}\bar{A}^{ij} \nonumber\\
    &-\frac{3}{2}\pi B\chi\psi^{4}\varphi^2T\phi^{-1}
    -\Big(\frac{3}{4}+\frac{5}{4B}\Big)\chi\psi^4m_\phi^2\varphi^2\phi\nonumber\\
    &-\frac{\chi}{4}\left(3+\frac{1}{B}-\frac{3}{4}\varphi^2 \right)f^{ij}(\partial_i\varphi)(\partial_j\varphi) \nonumber\\
    &-\frac{3}{4}\psi^{-1}\varphi f^{ij}(\psi\partial_i\chi-\chi\partial_i\psi)(\partial_j\varphi), 
\end{align}
\begin{align}\label{eq:elp3}
    \Delta\beta^{i}+\frac{1}{3}f^{ij}\partial_j(\partial_k\beta^k)=&\,\,16\pi \phi^{-1}\chi\psi^{-1}f^{ij}J_j \nonumber\\
    &-2\chi\psi^{-7}\bar{A}^{ij}(7\psi^{-1}\partial_j\psi-\chi^{-1}\partial_j\chi) \nonumber\\
    &-2\chi\varphi\psi^{-7}\bar{A}^{ij}\partial_j\varphi,
\end{align}
and
\begin{align}\label{eq:ellip_sca}
    \Delta\varphi=&\,\,2\pi B\psi^4\varphi\phi^{-1} T -\varphi f^{ij}(\partial_i\varphi)\partial_j\varphi \nonumber\\
    &-f^{ij}(\chi^{-1}\partial_i\chi+\psi^{-1}\partial_i\psi)(\partial_j\varphi) +m_\phi^2\psi^4\varphi\phi,
\end{align}
where $\Delta$ denotes the flat Laplacian, $\psi=W^{-1/2}$, $\chi=\alpha\psi$, $S=T_{ij}\gamma^{ij}$, and we used the definition
\begin{align}
    \bar{A}^{ij}=\psi^{10}\left(K^{ij}-\frac{1}{3}\gamma^{ij}K\right).
\end{align}
We also assumed that the ``momentum'' of the scalar field $\varphi$, denoted by $\Phi$, vanishes given that the scalar-radiation reaction time scale is much longer than the orbital time scale. From Eq.~\eqref{eq:ellip_sca} we see that the asymptotic value of the scalar field, $\varphi_0$, can be oscillatory (e.g., \cite{alby17}) or zero for stationary solutions. We adopt the latter case in the present work, i.e., $\varphi_0=0$. By modifying the elliptic equations \eqref{eq:elp1}--\eqref{eq:elp3} and introducing equation \eqref{eq:ellip_sca}, we generalise the public spectral code FUKA \cite{pape21} to this ST theory for generating the BNSs in quasi-equilibrium.

Note that, for large distances, FUKA uses a compactified domain to bring infinity to a finite numerical distance (this allows in particular to properly impose boundary conditions at infinity). Given the asymptotic exponential decay of the scalar field $\varphi$, its profile is better captured in such a domain if Eq.~\eqref{eq:ellip_sca} is rewritten in terms of an auxiliary scalar field $\xi=\varphi \cosh(m_\phi r)$, which gives
\begin{equation}\label{eq:ellip_sca_mod}
\begin{aligned}
    \Delta\xi=&m_\phi^2 \left[ 2 \cosh^{-2}(m_\phi r) + \psi^4 \phi - 1 \right] \xi \\
    &+\frac{2m_\phi \tanh(m_\phi r)}{r} \xi +2m_\phi \tanh(m_\phi r) \hat{r}^i\partial_i\xi \\
    &+2\pi B\psi^4\xi\phi^{-1}T-\cosh^{-2}(m_\phi r)\xi\Big[f^{ij}\partial_i\xi\partial_j\xi \\
    &-2m_\phi\xi\tanh(m_\phi r)\hat{r}^i\partial_i\xi+m_\phi^2\xi^2\tanh^{2}(m_\phi r)\Big] \\
    &-(\chi^{-1}\partial_i\chi+\psi^{-1}\partial_i\psi)  \Big[f^{ij}\partial_j\xi-m_\phi\xi\tanh(m_\phi r)\hat{r}^i\Big],
\end{aligned}
\end{equation}
where $\hat r^i$ is the unit radial vector. 
The first term in the right-hand side suggests a Helmholtzian nature of the equation, which, however, asymptotically reduces to a Laplacian one under the assumption of this paper that $\phi \rightarrow 1$ at $r \rightarrow \infty$.

\subsection{Spontaneous scalarization with massive fields}\label{sponsca}

\begin{figure}
    \centering
    \includegraphics[width=\columnwidth]{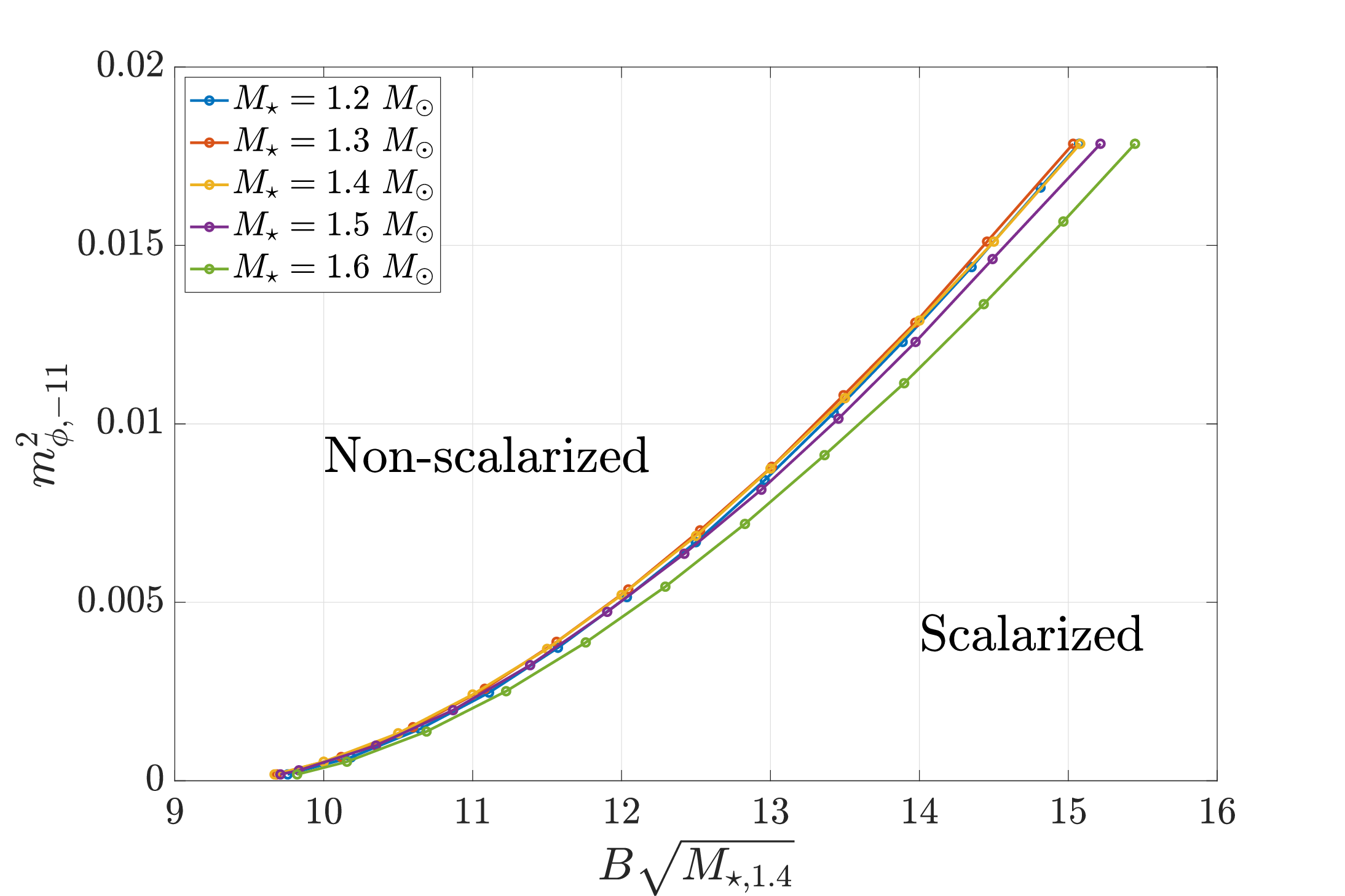}
    \caption{Boundaries of the scalarization projected on the $m_\phi^2-B\sqrt{\Ms{}_{,1.4}}$ plane for a variety of stellar masses, which separate the upper region where stars do not harbour a static scalar field from the lower region of scalarized NSs. Here the notations $\Ms{}_{,1.4}=\Ms/1.4\,\,M_\odot$ and $m_{\phi,-11}=m_\phi/10^{-11}$~eV are used, and the APR4 EOS is adopted.
    }
    \label{fig:mB}
\end{figure}

In isolated NSs and for a given coupling strength $B$, scalarization is triggered by tachyonic instability if the NS exceeds a threshold compactness determined by the theory parameters and the EOS. In particular, the conditions to be met for spontaneous scalarization in a spherical NS are approximately $k^2>0$ and $k\Rs\rightarrow\pi/2$ for $k^2=-(2\pi BT+m_\phi^2)$ \cite{kuro23}.
In the massless theory, the threshold is only weakly EOS-dependent for some coupling strength, given that $-T\approx\rho$ 
\cite{damo93,shib14,alta17,yagi21}. However, this universality is lost from the non-vanishing $m_\phi$ \cite{yaza16}. 
Instead of studying the EOS dependence of the threshold, we focus on a particular EOS (ARP4 \cite{APR}) and look at how the scalarization criterion is modified by $m_\phi$. 

In Fig.~\ref{fig:mB}, we trace out the marginally-scalarized configuration on the $m_\phi^2$--$B\sqrt{\Ms}$ plane where $\Ms$ denotes the mass of the NS (see Sec.~\ref{tensorM} for more details on defining stellar mass). We observe that the critical coupling strength $B$ for scalarization correlates approximately with the squared mass of the scalar field, and the relation depends only slightly on the specific stellar mass. For the considered EOS, we find the fitting formula
\begin{align}\label{eq:mB}
    \left( \frac{m_\phi}{1.6\times10^{-11}\,{\rm eV}}    \right)^2 \approx 1-2.52x+1.54x^2,
\end{align}
where
\begin{align}
    x=\left(\frac{B}{10}\right) \left(\frac{M_\star}{1.4\,M_\odot}\right)^{1/2}.
\end{align}
Therefore, for a given scalar mass, the critical coupling strength is approximately a function of $\Ms$. In particular, the critical coupling strength $B_{\rm crit}$ for massless ST theories is solved as
\begin{align}
    B_{\rm crit}^{m_\phi=0} \approx 9.6 \left( \frac{\Ms}{1.4\,M_\odot} \right)^{-1/2}.
\end{align}
We see also that the critical coupling strength increases monotonically with $m_\phi$ (i.e., $\partial B_{\rm crit}/\partial m_\phi>0$). This tendency continues up to the mass large enough to eliminate scalarization for any coupling strength \cite{rama16}. For NSs whose typical radius is $\sim10$~km, mass of $m_\phi\gtrsim 2 \times 10^{-11}$\,eV severely suppresses scalarization in NSs since the associated Compton length is shorter than the stellar size. We thus only consider masses smaller than this limit.

In addition, the presence of scalar hair provides extra supporting force, thus sustaining more matter for a given stellar mass (the meaning of stellar mass will be further clarified in Section \ref{tensorM}), i.e., the stellar rest mass
\begin{align}
    M_{\rm b}=\int \rho u^t\sqrt{-g}d^3x
\end{align}
is larger for stronger scalarization.
As an illustration, assuming $m_\phi=1.33 \times 10^{-11}$\,eV, EOS APR4, and $\Ms=1.35\,M_\odot$, one has $M_{\rm b}=1.5021\,M_\odot$ for $B=15.5$, while $M_{\rm b}$ increases by $0.015\,M_\odot$ for $B=17$.

\subsection{Current Constraints}

Pulsar-timing observations in NS-WD binaries \cite{frei12,shao17,ande19} or in galactic NS-NS binaries \cite{kram21} can constrain the parameters of ST theories based on scalar-wave emissivity (assuming $m_\phi \alt 10^{-19}$\,eV). In fact, the ST theory with a massless scalar field and a high coupling constant $B\agt 9$ (i.e., $\beta_\mathrm{DEF} \alt -4.5$) is ruled out by the network of pulsar systems \cite{chib22,zhao22}. However, a tiny value of $m_\phi > \lambdabar_\mathrm{gw}^{-1}$ (here $\lambdabar_\mathrm{gw}$ denotes the wavelength of scalar waves which is comparable to the gravitational wavelength) can account for the absence of scalar radiation and the reason is as follows. The propagation group speed of scalar waves ($v_g$) with the frequency $\omega_{\rm gw}$ can be approximately written as 
\begin{align}
    v_g=\sqrt{1-m_\phi^2\omega_{\rm gw}^{-2}}=(1+m_\phi^2\lambdabar_\mathrm{gw}^2)^{-1/2},
\end{align}
where we note that the relation between the wavelength and frequency is $\lambdabar_{\rm gw}=(\omega_{\rm gw}^2-m_\phi^2)^{-1/2}$.
This speed is much slower than the speed of light for $\lambdabar_\mathrm{gw} \gg \comp$, thus essentially prohibiting the scalar-wave emission.

Aside from the scalar-wave emissivity, the gravitational field around scalarized NSs can be appreciably different from that in GR within a few times of $\comp$ (see Fig.~\ref{fig:M_sc} below). Accordingly, the orbital motion around the scalarized NS should be modified for orbital separations comparable with $\comp$. Such modification is, however, not seen in the observations. A small value of the mass $m_\phi \gg 1/a\,(\sim 10^{-16}$\,eV for observed NS-WD systems) is then necessary to circumvent the current observational constraint if the NSs are scalarized (e.g., \cite{rama16}). We note that this mass range can also accommodate what is observed from the triple system PSR J0337+1715 \cite{seym20} (see Fig.~2 therein).
In addition, simultaneous mass-radius measurements by monitoring rotating hot spot patterns of pulsars can also probe the theory parameters \cite{tuna22,demi23}, while the constraints obtained in this way are currently weaker than the aforementioned ones.

The tensorial gravitational waveforms observed for a BNS can constrain the theory by measuring the scalar-radiation-induced phase shift \cite{quar23}. For the specific event GW170817, the observation does not support significant scalar effects in inspiral stages \cite{zhao19,meht23}, while the induced/dynamical scalarization in late-inspiral-to-merger phase remains unconstrained due to the insufficient sensitivity to the lase inspiral waveform. An upper limit of $B\lesssim9-9.4$ is thus suggested for \emph{massless} scalar field if the two NSs are slowly rotating (cf.~Fig.~13 in \cite{zhao19}; we note again that their parameter is $\beta=-B/2$). This constraint substantially prevents spontaneous scalarization in NSs. In order to revive the existence of scalarized NSs, the Compton length-scale has to be much smaller than the constraint from the pulsar systems, because the orbital separation of inspiraling BNSs in the range of GW observations is quite small, within $\sim 20$--200\,km. However, the scalar effects in this regime is not trivial, so that the present numerical work is required; see Sec.~\ref{secIV} for more details.

Although much less stringent, the gravitational phenomena in the solar system (e.g., Shapiro time delay measured by Cassini tracking) put constraints on the scalar mass $m_\phi\gtrsim 10^{-17}$\,eV \cite{peri10,alsi12}. Possible constraints on the massive theories may also be placed by extreme mass-ratio inspirals (EMRIs) where superradiance modifies orbital dynamics \cite{brit15}, e.g., with the presence of floating orbits on resonance `islands' \cite{yune12}, thus leading to phase shifts in gravitational waveforms (much similar to the ramification of non-Kerr black hole spacetimes \cite{dest20,dest21}). However, it has recently been pointed out that the scalar imprint in the waveforms may be indistinguishable from GR waveform baselines for $m_\phi\lesssim 4 \times 10^{-12}$\,eV \cite{bara23}. 

\section{Virial theorem, tensor mass, and asymptotic behavior of the geometry}\label{secIII}

\begin{figure*}
    \centering
    \includegraphics[width=\columnwidth]{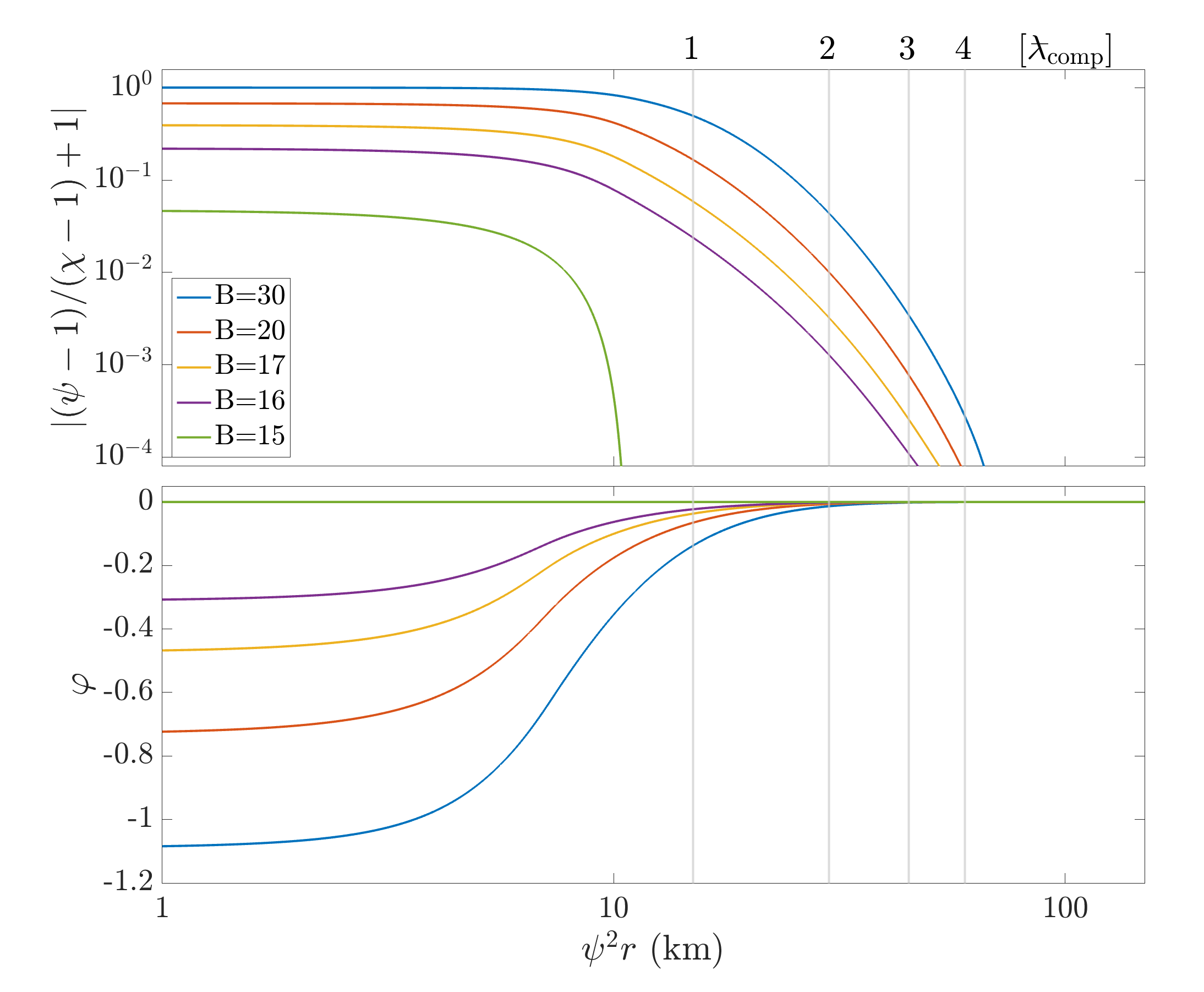}
    \includegraphics[width=\columnwidth]{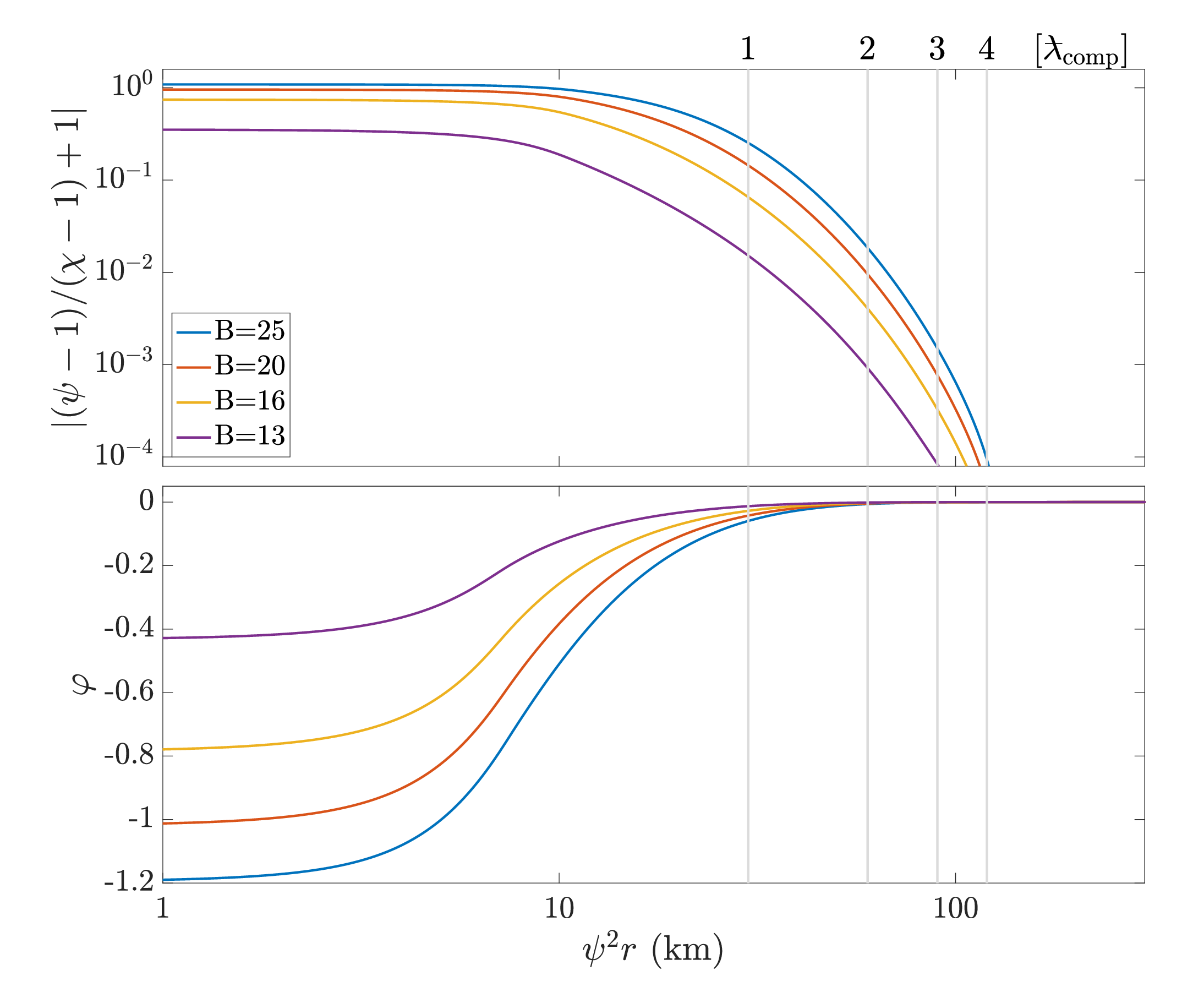}
    \caption{Relevant properties of isolated NSs in ST theories: Deviation between $(\psi-1)$ and $(\chi-1)$ as a function of the areal radius $\psi^2r$ (top) and the profile of the scalar field (bottom) for NSs with $\Ms=1.35\,M_\odot$. Theories with $\comp=15$\,km ($m_\phi=1.33\times10^{-11}$\,eV; left) and $\comp=30$\,km ($m_\phi=6.65\times10^{-12}$\,eV; right) are considered. For each scalar mass, four coupling strengths are adopted and listed in the legend. Note that the NS for $B=15$ and $\comp=15$\,km is not scalarized, and thus, the geometry is the same as in GR. Vertical lines mark the first four times of the associated Compton length-scale. The stellar radius (areal radius) for this model is $\approx 11.1$\,km. 
    }
\label{fig:M_sc}
\end{figure*}
In the present article, we assume the conformally flat (Isenberg-Wilson-Mathews~\cite{isen08, wils89}) approximation (see \cite{Friedman:2001pf,shib04,uryu06,uryu09} for a construction without this approximation), helical symmetry, and maximal slicing (i.e., $K=0$) for the spacetime. The quasi-equilibrium states in this formalism satisfies the viral relation~\cite{bona94,Friedman:2001pf,shib04,shib13}. Thus, we will validate the numerical solutions of the quasi-equilibria by the virial theorem, which is described for massive ST theory in Sec.~\ref{virial}. We then define the tensor mass in Sec.~\ref{tensorM}, which characterizes the physical mass of the system.

\subsection{Virial theorem}\label{virial}

Given that the asymptotic behavior of the scalar field in the Einstein frame reads 
\begin{align}\label{eq:asymphi}
    \bar{\varphi} = \bar{\varphi}_0 + \frac{M_{\bar{\varphi}}}{r}e^{-m_\phi r}+O(r^{-2}),
\end{align}
we have the following relations for $r\rightarrow\infty$, 
\begin{align}
    \varphi=\varphi_0 + \frac{ \sqrt{B} M_{\bar{\varphi}} }{r} e^{-m_\phi r}+ O(r^{-2}).
\end{align}
Since $\varphi$ approaches $\varphi_0$ exponentially at $r\rightarrow\infty$, the scalar charges $\sqrt{B}M_{\bar \varphi}$ does not contribute to the mass in the system. Thus, the virial relation is written in the same form as in GR (cf.~\cite{shib13})
\begin{align}
    \kom=M_{\rm ADM}, \label{eq:virial}
\end{align}
where $M_\mathrm{ADM}$ is the Arnowitt–Deser–Misner (ADM) mass and $M_\mathrm{K}$ denotes the Komar mass defined by
\begin{align}
    \kom=-\frac{1}{4\pi\phi_0}\oint_\infty dS_a n_b \phi\nabla^a\xi^b,
\end{align}
where we have assumed the existence of a timelike Killing vector $\xi^a$ fulfilling $n_a\xi^a=-\alpha$.

\subsection{Tensor Mass}\label{tensorM}

As the ADM mass in the Einstein frame decreases monotonically when GWs propagate away and is positively defined \cite{lee74,sche95,sche95II}, we refer it to the mass of a given system following, e.g., \cite{done13,shib13}, and define it as the tensor mass $M_{\rm T}$ to be distinguishable from the ADM mass in the Jordan frame ($M_{\rm ADM}$). As a specific example, the stellar mass refers to the tensor mass of a NS, i.e., $\Ms=M_{\rm T}$. In the massless ST theories, the tensor mass is written as the sum of the ADM mass and scalar charge~\cite{lee74}. As mentioned in Sec.~\ref{virial}, the scalar charge does not contribute to the mass of the system in the massive ST theories. Thus, we simply have $M_\mathrm{T}=M_\mathrm{ADM}$. If the virial theorem is satisfied, the tensor mass is also equal to the Komar mass.

\subsection{Asymptotic behavior of the geometry}

In GR, the asymptotic behavior of $\psi$ and $\chi$ at a large distance in isotropic coordinates is described as (e.g., \cite{Friedman:2001pf})
\begin{eqnarray}
\psi &=& 1 + {M_\mathrm{ADM} \over 2r} + O(r^{-2}),\\
\chi &=& 1 -{2 M_\mathrm{K}-M_\mathrm{ADM} \over 2r} + O(r^{-2}). 
\end{eqnarray}
Thus, the equality
\begin{equation}
(\psi-1)r=-(\chi-1)r, \label{eq33}
\end{equation}
holds at $r \rightarrow \infty$, if the virial relation is satisfied. For spherical stars in equilibrium, this relation is satisfied for the entire region outside the stellar surface, $r=\Rs$, because of the presence of Birkhoff's theorem in GR \cite{jebs05,birk23}. 

By contrast, Eq.~\eqref{eq33} is satisfied only at $r \rightarrow \infty$ in ST theories because the scalar clouds contribute to $\psi$ and $\chi$ in a different way. The deviation from the equality of Eq.~\eqref{eq33} outside the star is considered as a manifestation of ST theories. In particular, we plot in Fig.~\ref{fig:M_sc} the violation of the equality of $\psi-1=1-\chi$ (upper panels) and the profile of $\varphi$ (lower panels) for spherical NS models with $\Ms=1.35\,M_\odot$. Two scalar masses are considered with the associated Compton length-scale being $\comp \simeq 15$\,km (left) and 30\,km (right). By picking several values of $B$ for each value of $m_\phi$, we consider NSs scalarized to different extents. We see that the equality \eqref{eq33} holds for $r \gg \comp$, while the deviation can be $\agt 10^{-2}\%$ for $\psi^2r \alt 4\comp$ where the amplitude of the scalar field is appreciably non-zero. This clearly indicates that the presence of the scalar cloud can appreciably modify the binary motion if the orbital separation is smaller than a few times of $\comp$. 

It is also found that for larger values of $B$, the maximum value of $|\varphi|$ is larger, and as a result, the region, in which the equality of Eq.~(\ref{eq33}), is breached is wider. Thus, for larger values of $B$, the scalar could modifies the binary motion from a larger distance (see Sec.~\ref{sec4.1}).

\section{Binary neutron stars in quasi-equilibria}\label{secIV}

\begin{figure}
    \centering
    \includegraphics[width=\columnwidth]{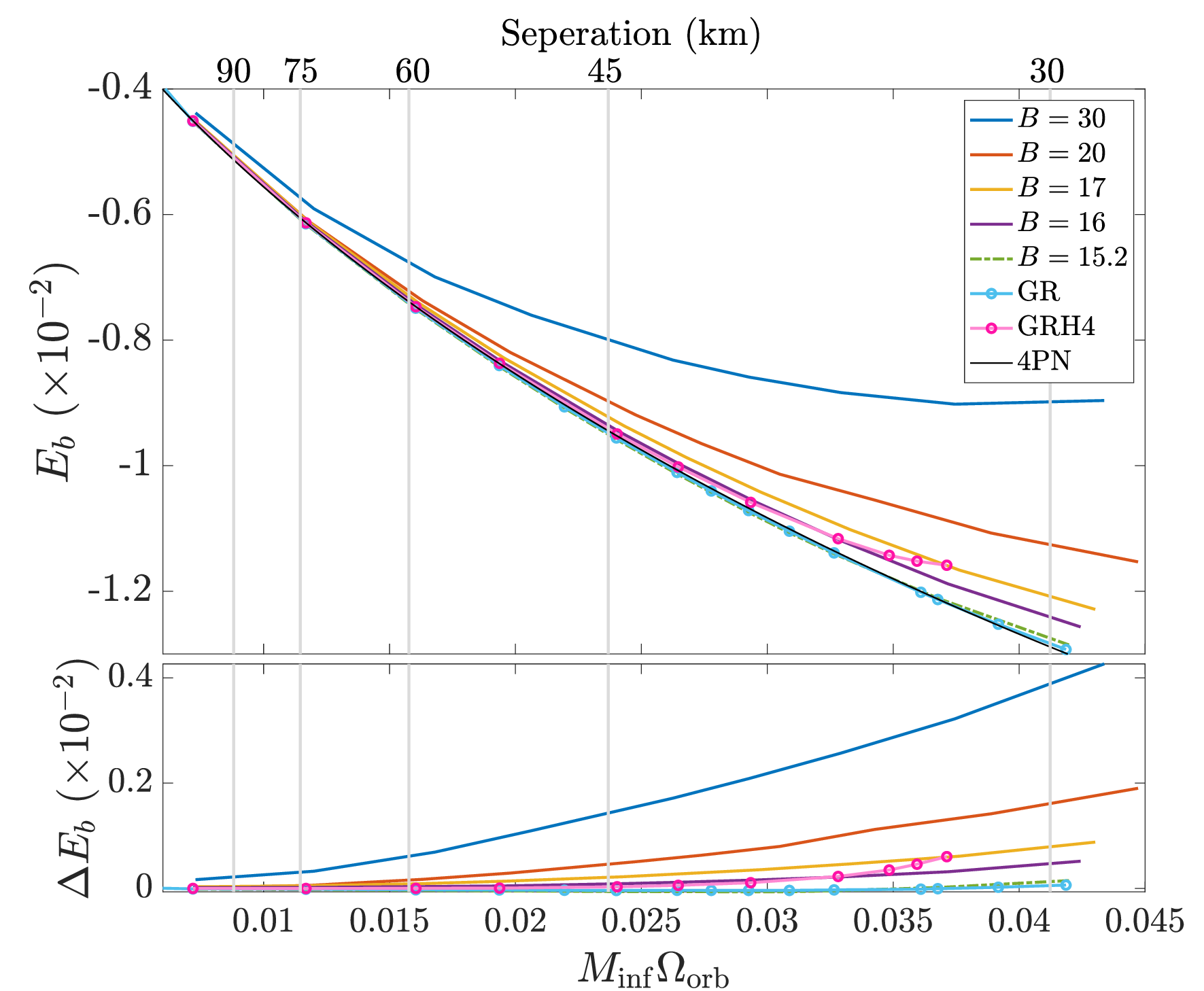}
    \includegraphics[width=\columnwidth]{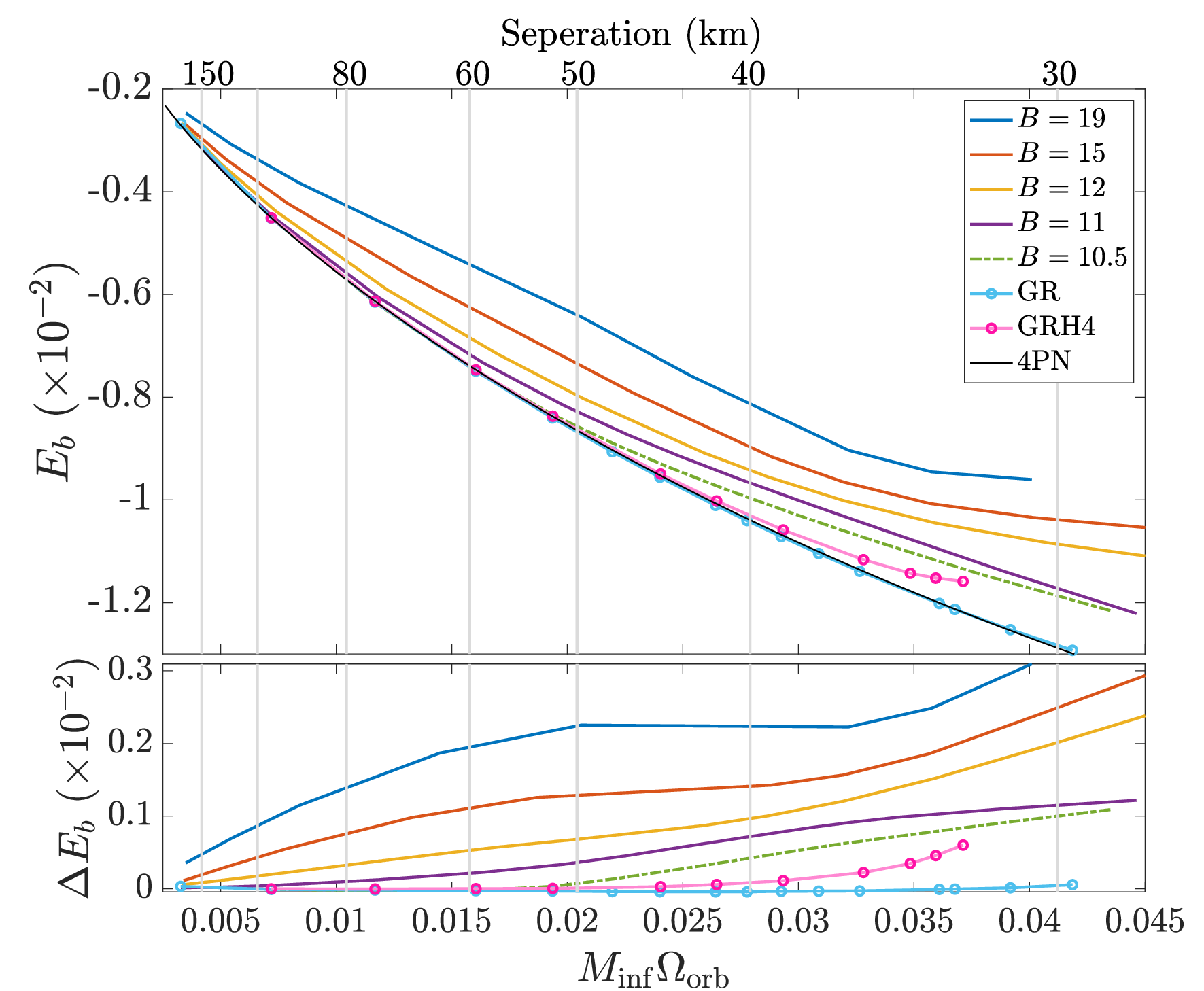}
    \caption{Quasi-equilibrium sequences for symmetric binaries with each NS having $1.35\,M_\odot$. The binding energy is plotted as a function of orbital frequency in the top, while the deviation of various theories from the 4PN analytic estimates of GR is shown in the bottom. Two EOS have been adopted for GR sequences, viz.~APR4 (blue-circle) and H4 (pink-circle), while two scalar masses, $m_\phi=1.33\times10^{-11}$~eV (top; $\comp\simeq15$~km) and $m_\phi=4\times10^{-12}$~eV (bottom; $\comp\simeq50$~km), are considered for ST sequences.
    Depending on different ST parameters, scenarios of dynamical enhancement of the scalarization (colorful solid) and dynamical scalarization (dash-dotted) manifest. The vertical gray lines relate the binary separation and $M_{\rm inf}\Oo$ based on the GR sequence.
    }
    \label{fig:seq}
\end{figure}

The major purpose of this paper is to clarify in which cases the effect of the scalarization of NSs can be identified by observing GWs from inspiraling BNSs. 
Given that the current GW detectors are able to detect signals for $f \approx 20$--$10^3$~Hz, where the separation between the members of a BNS is less than $\sim 200$\,km (for NS masses of $\sim 1.4M_\odot$), the scalar mass of interest will then be 
\begin{align}
    m_\phi \ge 1 \times 10^{-12}\,\mathrm{eV},
\end{align}
associated with Compton length scales of $\le 200$\,km. We consider $m_\phi=4\times10^{-12}\,{\rm eV}$ ($\comp \approx 50$\,km) and $m_\phi=1.33\times10^{-11}\,{\rm eV}$ ($\comp \approx 15$\,km) as two canonical cases to demonstrate the role played by the scalar mass, as well as coupling strength, in the last several orbits of BNSs. To model the hydrodynamical equilibria of NSs, we adopted the piecewise-polytropic approximated EOS APR4 \cite{read09}. The details of our implementation are essentially the same as those in \cite{tani10}, and thus we will not repeat them here.

Denoting the tensor masses of the two NSs when they were in isolation as $M_{\star,1}$ and $M_{\star,2}$,  the total mass $M_{\rm inf}=M_{\star,1}+M_{\star,2}$ is kept constant along each binary sequence. In this work, we choose $M_{\rm inf}=2.7\,M_\odot$, while consider two values for the mass ratio, viz.~$q=M_{\star,2}/M_{\star,1}=1$ and $0.8$. Each quasi-equilibrium state on a particular sequence is characterized by a dimensionless orbital angular velocity $M_\mathrm{inf}\Omega_\mathrm{orb}$ and the orbital binding energy defined by
\begin{align}
    E_{\rm b}=\frac{M_{\rm T}-M_{\rm inf}}{M_{\rm inf}}.
\end{align}
We compare the curves of $E_\mathrm{b}$ as a function of $M_\mathrm{inf}\Omega_\mathrm{orb}$ with that in GR and identify the effect of the scalar field. Specifically, we will show that the scalar-related dynamical response in the late time can noticeably expedite the merger (Sec.~\ref{cycles}), while the orbital frequency at the last orbit increases only slightly compared to the GR value \cite{jain23,juli22} (see also Section \ref{shedding}).

The quality of the constructed configurations is examined by checking the violation of Eq.~\eqref{eq:virial}, i.e.,
\begin{align}\label{eq:vir}
    {\cal E}_{\rm virial}=\frac{|\kom-M_{\rm ADM}|}{M_{\rm ADM}},
\end{align}
which has been found to be less than $0.06\%$ for our results. In addition, we evolved some of the obtained quasi-equilibrium states with our numerical code (developed from the previous code~\cite{shib14}) for a few orbits to validate our ID solver. We confirmed that the BNSs have quasi-circular orbits with a small eccentricity of $10^{-2}$, which is approximately the same magnitude as that in \cite{tani14}.

\subsection{Quasi-equilibrium sequences}\label{sec4.1}

\begin{figure}
    \centering
    \includegraphics[width=\columnwidth]{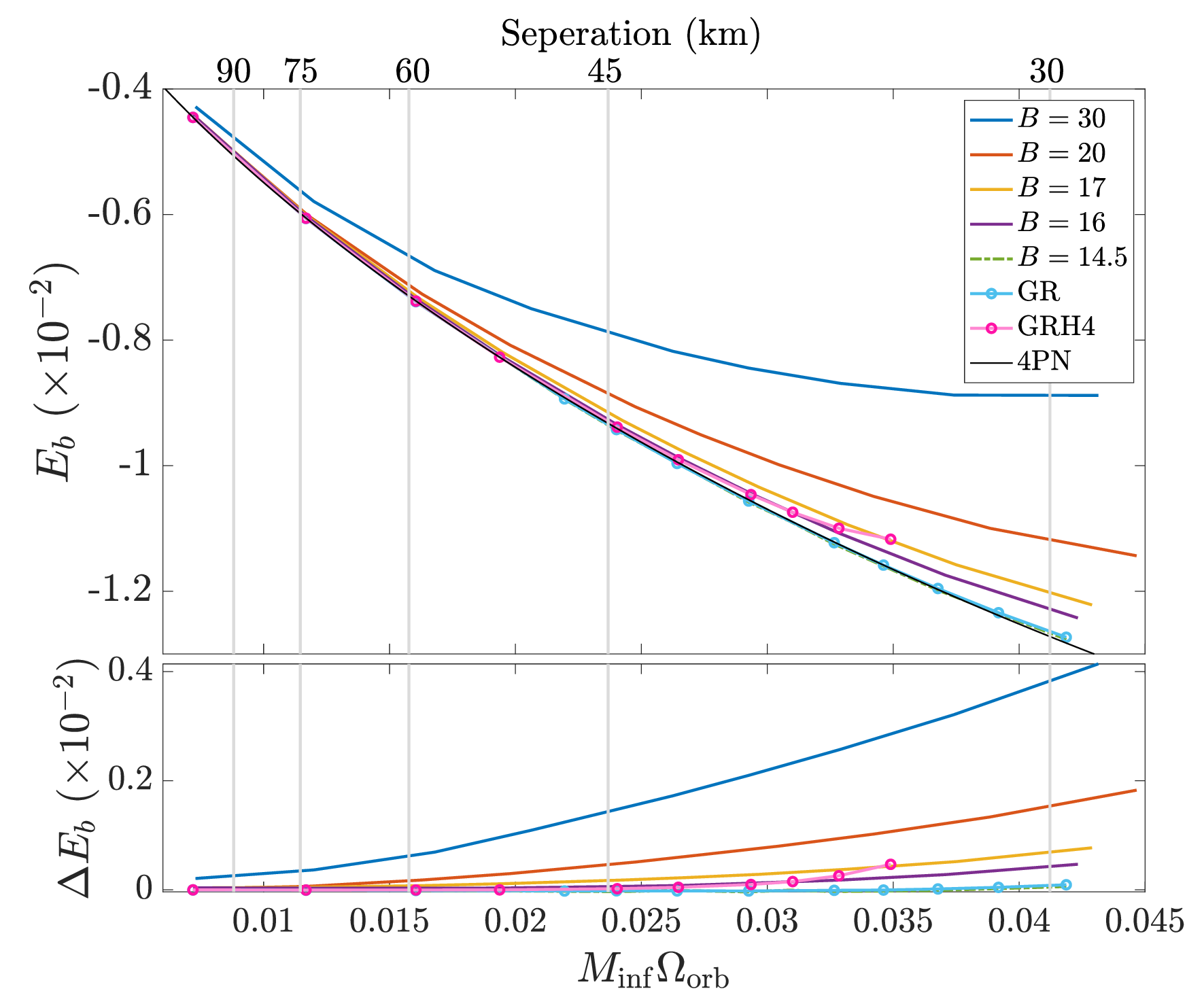}
    \includegraphics[width=\columnwidth]{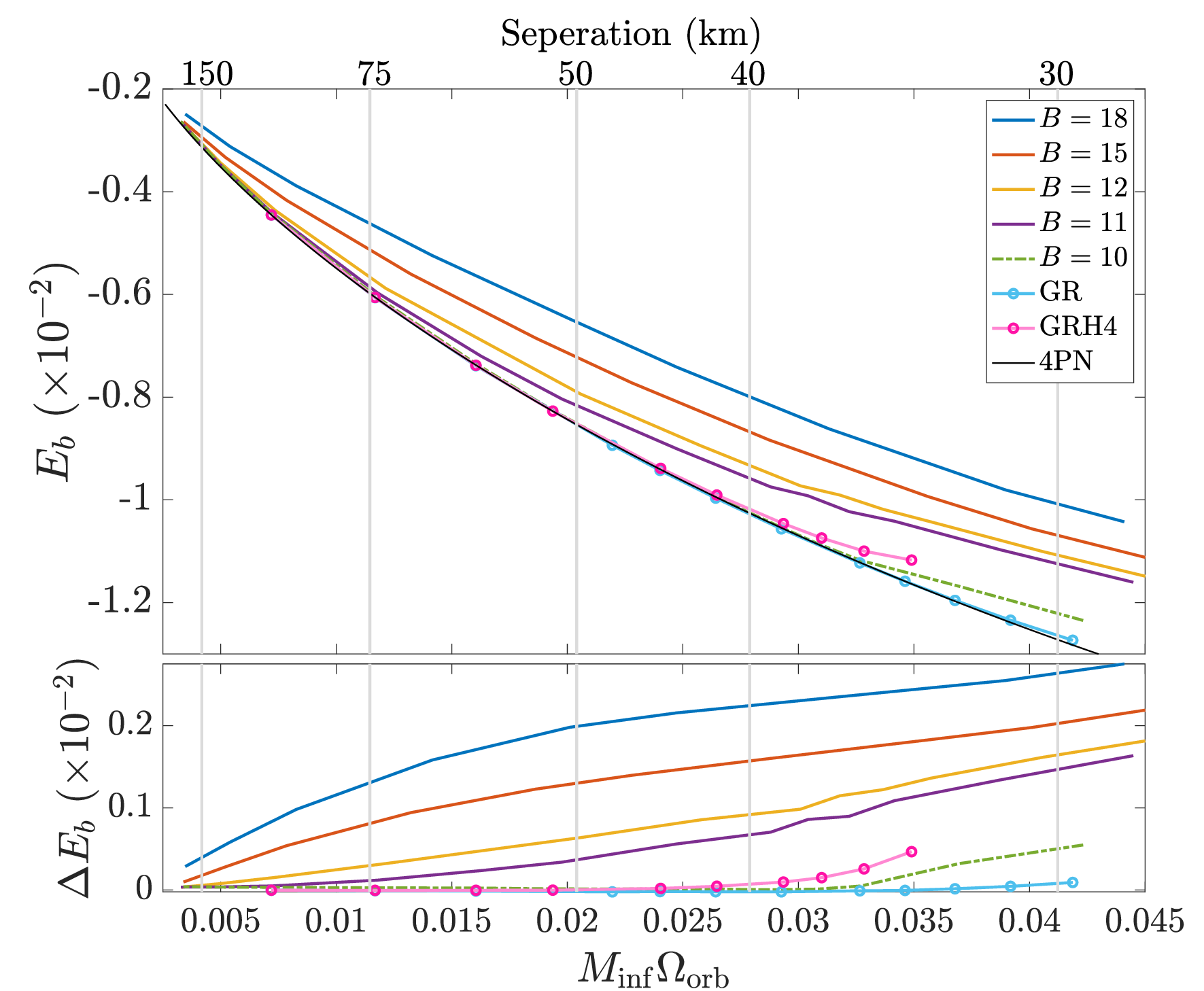}
    \caption{Same as Fig.~\ref{fig:seq}, but for asymmetric binaries with $1.5\,M_\odot+1.2\,M_\odot$.
    }
    \label{fig:seqII}
\end{figure}

\begin{figure}
    \centering
    \includegraphics[width=\columnwidth]{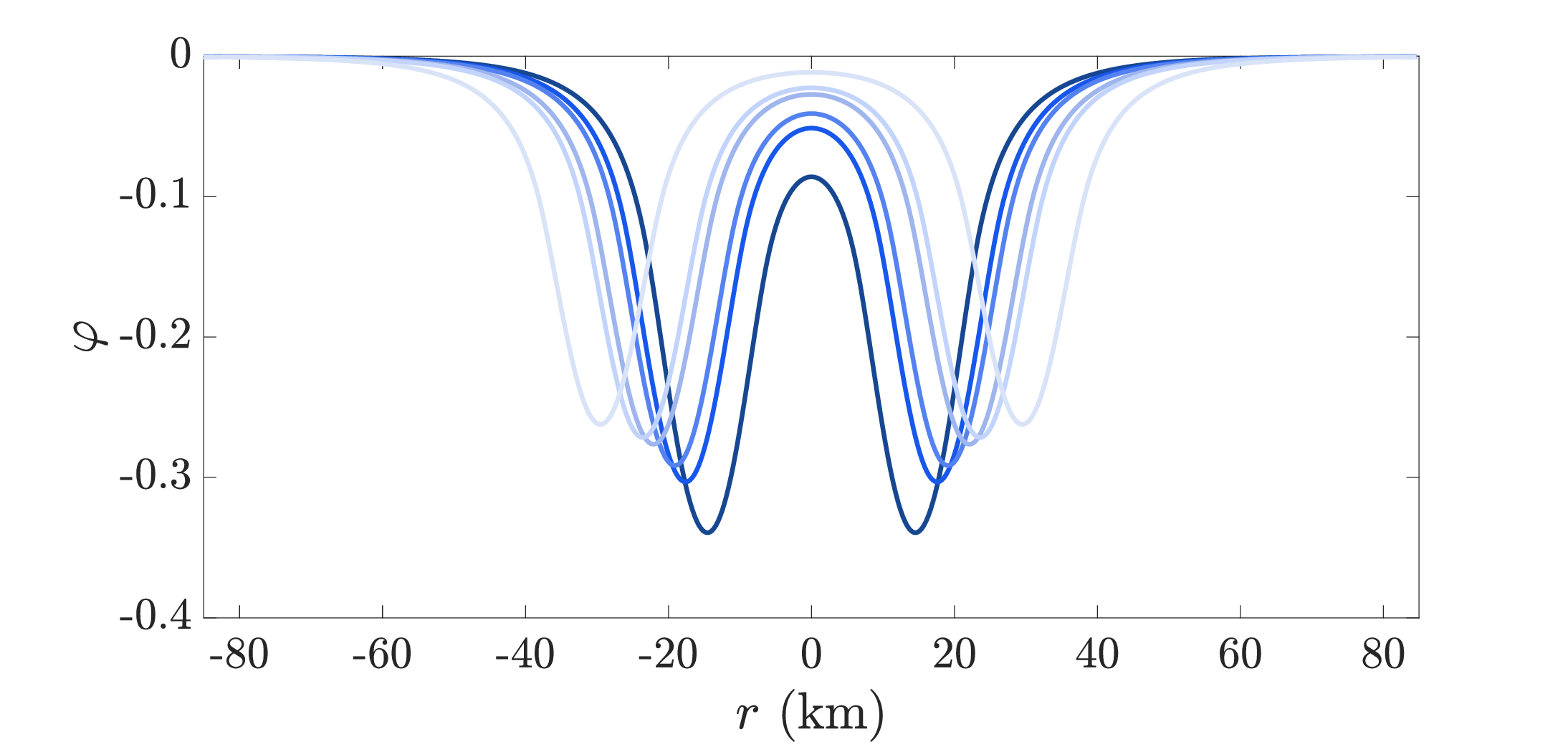}
    \includegraphics[width=\columnwidth]{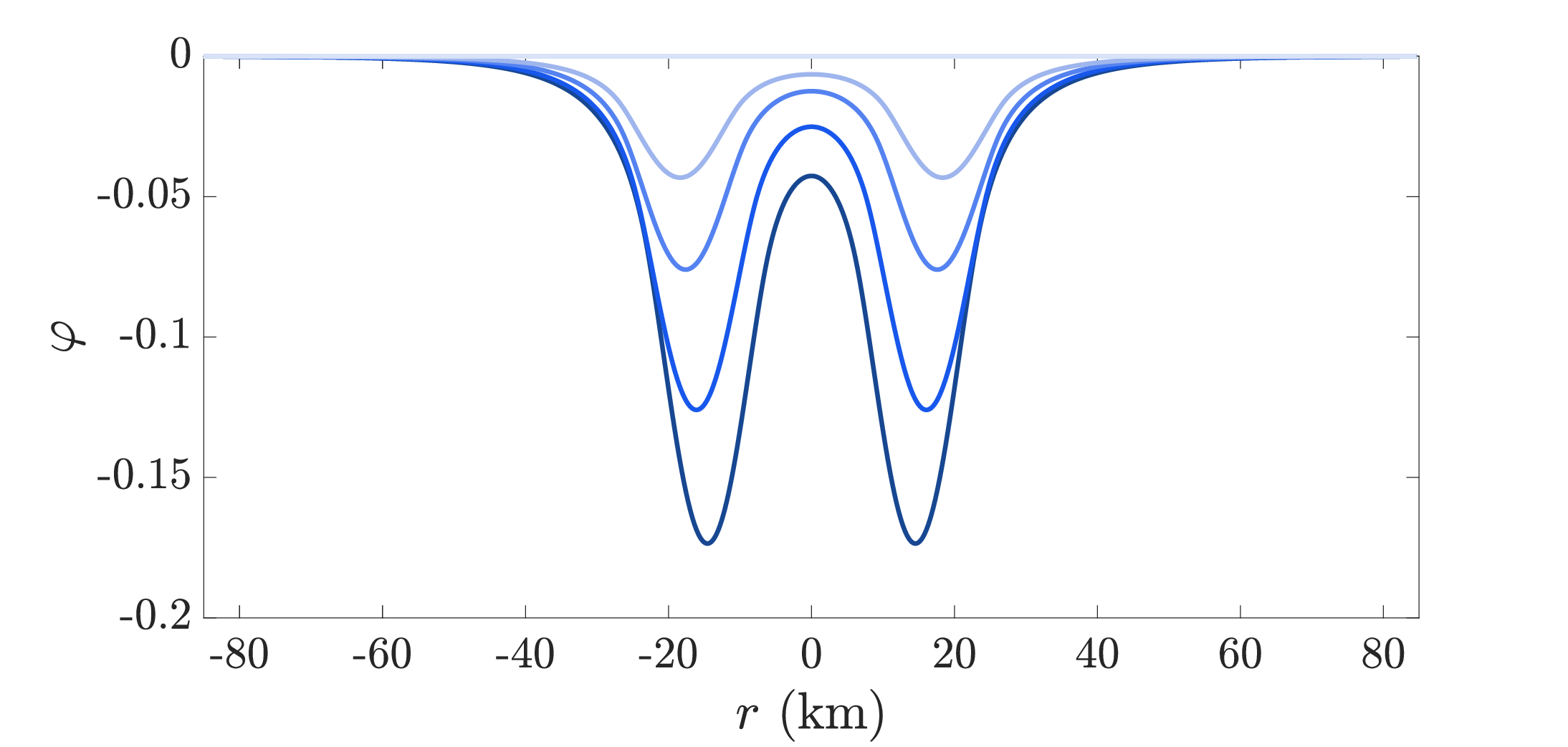}
    \caption{Radial profile of the scalar field for equal-mass binaries at different stages, undergoing enhancement of scalarization (top) and dynamical scalarization (bottom). The color is darker for closer separation, and the coupling strengthes are set to $B=15.8$ (top) and $B=15.2$ (bottom), respectively. The scalar mass here is assumed to be $m_\phi=1.33\times10^{-11}\,{\rm eV}$, and APR4 EOS is adopted.
    }
    \label{fig:enhance}
\end{figure}

In Figs.~\ref{fig:seq} and \ref{fig:seqII}, we plot the binding energy of binaries as a function of their orbital frequency. To represent the evolution track of a BNS, as least to the leading order, the rest mass of binaries is constant along each sequence \cite{uryu00}, while we note that it may vary from one sequence to another depending on $B$ and $m_\phi$ (see Sec.~\ref{sponsca}). The virial violation \eqref{eq:vir} for the constructed binaries is at most $0.06$\%, i.e., much smaller than the absolute value of the orbital binding energy. In both figures, we also show the GR curve (solid-circle) constructed by the original FUKA library \cite{pape21} for the EOS APR4 (light blue) and H4 (pink), and 4th order post-Newtonian (PN) approximation \cite{bern18,blan19} to clarify the scalar imprints. The deviation of the numerically constructed sequences from the 4PN prediction is denoted by $\Delta E_{\rm b}$ (bottom panels). Estimating the adiabatic tidal contribution by the difference between the GR sequence and the 4PN estimates, we see that scalar effects are similar to the enhanced tidal response for equal-mass binaries, so that systems with a soft EOS in ST theory could accidentally be identified as GR binaries with a \emph{stiffer} EOS (see also below). 

For each considered scalar mass, we choose 4 coupling strengthes that admit spontaneous scalarization (solid), as well as one slightly below the critical value (dashed-dotted). The former leads to the scenario of dynamical enhancement of the scalarization at a close orbit, resulting from the scalar-cloud interaction (see the upper panel of Fig.~\ref{fig:enhance}), while for the latter, the scenario is similar to the so-called dynamical scalarization (see the lower panel of Fig.~\ref{fig:enhance}), although the mechanism of the scalar-field enhancement is identical for both cases. The dynamical scalarization takes place for an orbital separation of $a \alt 1.7\comp$, slightly outside the Compton length scale, while the dynamical enhancement of the scalarization can do for more distant orbits of $a\alt 3$--$6\comp$ mainly contingent on the scalar mass. This enhancement starts at more distant orbits for larger values of $B$. The reason for this enhancement of the scalar fields outside the Compton length-scale is that even though the scalar field amplitude of one star decays exponentially outside that scale, it still has an appreciable value along the line connecting to its companion when the orbital separation is close enough. The same applies to the scalar field in the companion. The interaction between the tails of the scalar field induces a phenomenon similar to dynamical scalarization, leading to the enhancement of the scalar cloud around each NS. We note that for lower values of $B$ with which the maximum amplitude of the scalar field is low, i.e., $\varphi \alt 10^{-2}$, the enhancement of the scalar amplitude does not appreciably take place.

It is worth noting that BNSs follow the same evolution track as in GR even if spontaneously scalarized NS is present when $a\gtrsim 3$--$6\comp$ for the cases considered here, viz.~$m_\phi=1.33\times10^{-11}$ eV (top panels) and $m_\phi=4\times10^{-12}$ eV (bottom panels) cases. This critical distance within which the scalar imprint reveals matches well with the size of the scalar cloud of an isolated NS (Fig.~\ref{fig:M_sc}).
During this epoch [Stage (I) defined in Sec.~\ref{secI}], the scalar-wave emission is also negligible because the relation $\lambdabar_\mathrm{gw} > \comp$ is satisfied, and therefore, the ST theory is likely indistinguishable from GR. This indicates that for $\comp \alt 10$\,km (i.e., $m_\phi \agt 2 \times 10^{-11}$\,eV), the orbital evolution in this ST theory agrees with that in GR. 

As the binary separation shrinks to $a\lesssim 3$--$6\comp$ while $\lambdabar_{\rm gw}$ is still larger than $\comp$ [Stage (II)], we can observe the bifurcation of scalarized sequences from GR ones in both figures, though the scalar-emission is expected to be highly suppressed by the scalar mass. This $m_\phi$-induced suppression will however be eventually avoided when the binary evolves to Stage (III). The difference between (II) and (III) cannot be seen in quasi-equilibrium sequences since the radiation is approximately ignored in construction. In a future work, we will revisit this aspect.

\begin{figure}
    \centering
    \includegraphics[width=\columnwidth]{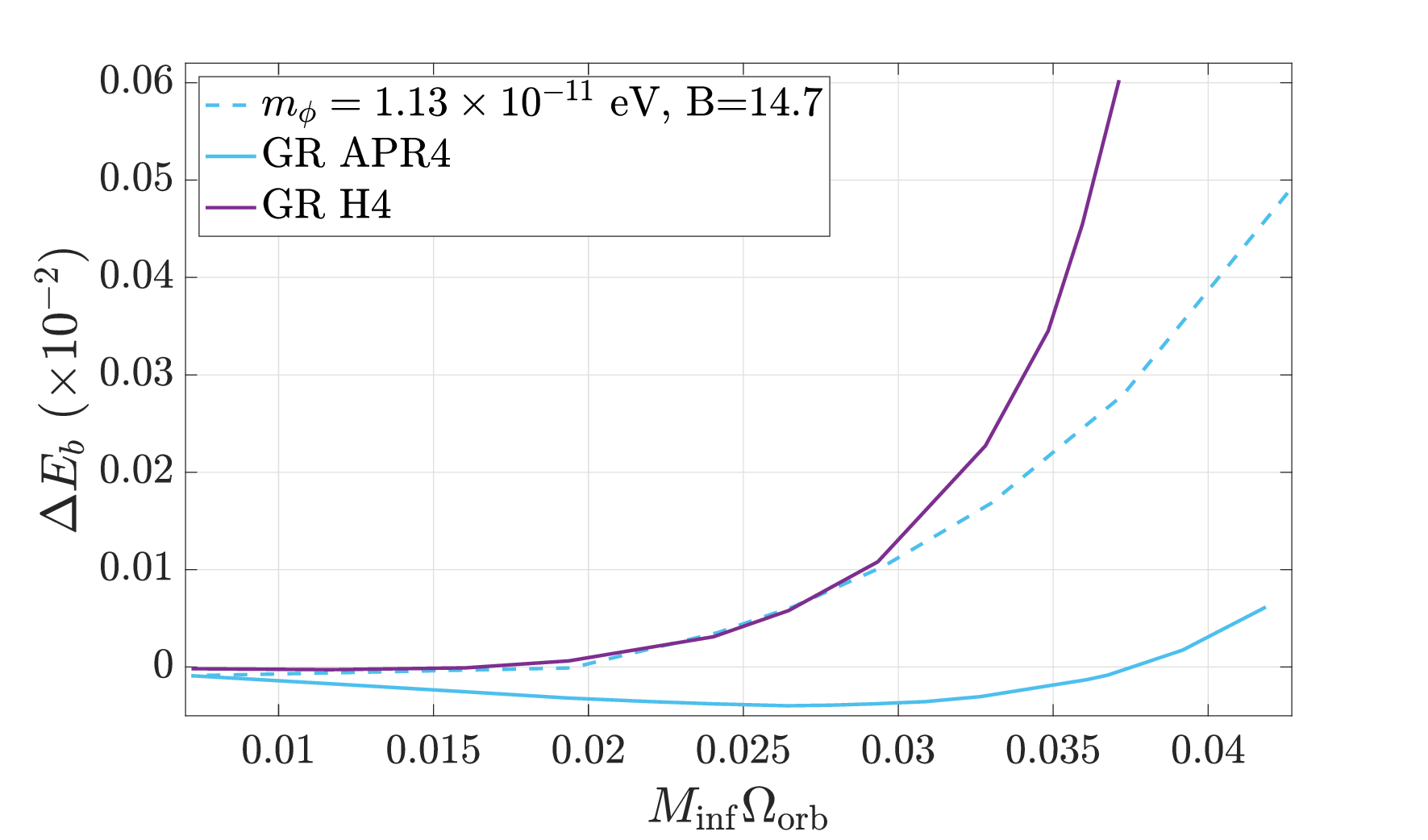}
    \caption{Deviation from 4PN approximant in the binding energy as a function of orbital frequency. Two EOSs, APR4 (blue curves) and H4 (purple curve), are employed. Einstein's gravity is assumed for both EOSs (solid curves), on top of which the curve of one specific ST theory with EOS APR4 is overplotted (dash-dot curve).
    }
    \label{fig:cmpr}
\end{figure}

A word of caution is appropriate here. The curves of $E_\mathrm{b}$ for a non-zero mass $m_\phi$ cases are similar to those in GR assuming a stiffer NS EOS, where the NS radius (i.e. tidal deformability) is high enough (see e.g., \cite{tani10}). For example, we plot in Fig.~\ref{fig:cmpr} the deviation from 4PN binding energy for a sequence of a particular ST theory with the EOS APR4, and for a GR sequence with the EOS H4, for which the tidal deformability is about 3.5 times larger than that for EOS APR4~\cite{Hotokezaka:2015xka}. We see that the two curves coincide when $M_{\inf}\Oo < 0.03$, indicating that the effect of the scalar-field interaction entangles with that of the NS EOS until late-inspiral. On the other hand, the curves of $E_\mathrm{b}$ for $m_\phi=4\times10^{-12}$ eV cannot be reproduced by the NS EOS effect because the deviation from the GR curve sets in at a distant orbit. An approximate estimate taking into account the previous GR studies (e.g., \cite{tani10}) gives that the tidal effect of the NS is appreciable only for $M_\mathrm{inf}\Omega_\mathrm{orb}\agt 0.02$ (i.e., an orbital separation of $\sim 50$\,km) for a NS with a radius of $\sim 15$\,km. Therefore, if $\comp \agt 20$\,km, the scalar-field interaction effect may be distinguished from the NS EOS effect assuming that the NS radius is less than 14\,km~\cite{abbo17prl,abbo19prx}. This suggests that by observing GWs from BNSs, the mass of the scalar field could be bounded from below for a hypothetically high value of $B$.

\subsection{Cycles in gravitational waveform}\label{cycles}

\begin{table}
    \centering
    \begin{tabular}{p{0.18\textwidth}|
				>{\centering}p{0.18\textwidth}
				>{\centering\arraybackslash}p{0.1\textwidth}}
    \hline
    \hline \\[-.8em]
        Binary components    & $(m_\phi,\,B)$ & ${\cal N}$ \\   \\[-.8em]      
    \hline\\[-.8em]
    \multirow{10}{*}{$1.35M_\odot+1.35M_\odot$} 
        & (0.03, 10.5)               & 25.66 \\
        & (0.03,\,\,\,\,\,\,11)      & 24.62 \\
        & (0.03,\,\,\,\,\,\,12)      & 23.33 \\
        & (0.03,\,\,\,\,\,\,15)      & 21.86 \\
        & (0.03,\,\,\,\,\,\,19)      & 19.80 \\
        \\[-.8em]\cline{2-3}\\[-.8em]
        & (0.1\,\,,\,\,15.2)         & 27.27 \\
        & (0.1\,\,,\,\,\,\,\,\,16)   & 26.65 \\
        & (0.1\,\,,\,\,\,\,\,\,17)   & 25.92 \\
        & (0.1\,\,,\,\,\,\,\,\,20)   & 22.13 \\
        & (0.1\,\,,\,\,\,\,\,\,30)   & 21.13 \\[-.8em]\\
    \hline \\[-.8em]
    \multirow{10}{*}{$1.5M_\odot+1.2M_\odot$}
        & (0.03,\,\,\,\,\,\,10)      & 27.46 \\
        & (0.03,\,\,\,\,\,\,11)      & 24.60 \\
        & (0.03,\,\,\,\,\,\,12)      & 23.74 \\
        & (0.03,\,\,\,\,\,\,15)      & 22.34 \\
        & (0.03,\,\,\,\,\,\,18)      & 20.60\\
        \\[-.8em]\cline{2-3}\\[-.8em]
        & (0.1\,\,,\,\,14.5)         & 27.71 \\
        & (0.1\,\,,\,\,\,\,\,\,16)   & 27.34 \\
        & (0.1\,\,,\,\,\,\,\,\,17)   & 26.64 \\
        & (0.1\,\,,\,\,\,\,\,\,20)   & 24.60 \\
        & (0.1\,\,,\,\,\,\,\,\,29)   & 20.84 \\[-.8em]\\
    \hline
    \hline
    \end{tabular}
    \caption{Number of cycles when the binary evolves during $f_{\rm gw}=240$--$957$~Hz for a variety of ST parameters. In this table we present the dimensionless scalar mass with a note that $m_\phi=0.1=1.33\times10^{-11}$~eV.}.
    \label{tab:cycles}
\end{table}

\begin{figure}
    \centering
    \includegraphics[width=\columnwidth]{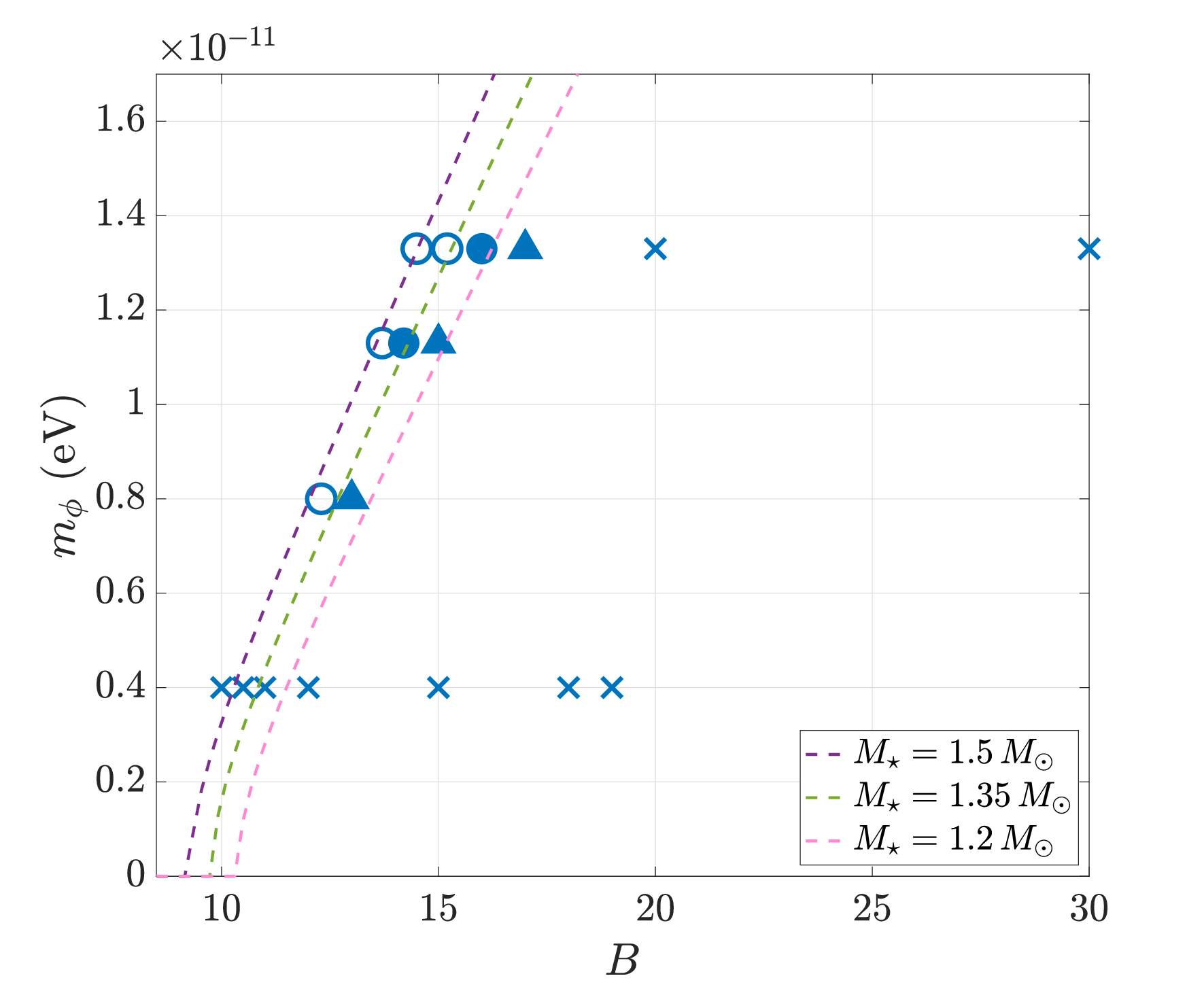}
    \caption{Parameter space of the considered massive ST theory. Relation \eqref{eq:mB} for stellar masses of $1.5\,M_\odot$ (purple), $1.35\,M_\odot$ (green), and $1.2\,M_\odot$ (pink) are plotted as dashed lines. The markers present the viability of the corresponded ST theory  after GW170817 especially for binaries with spontaneously scalarized NSs (filled markers). Specifically, circles (crosses) denote (un)acceptable parameters concerning with the two chosen binary configurations, while triangle marks the theory only allowed by the $1.5\,M_\odot+1.2\,M_\odot$ binary.
    }
    \label{fig:matome}
\end{figure}

\begin{figure*}
    \centering
    \includegraphics[width=\columnwidth]{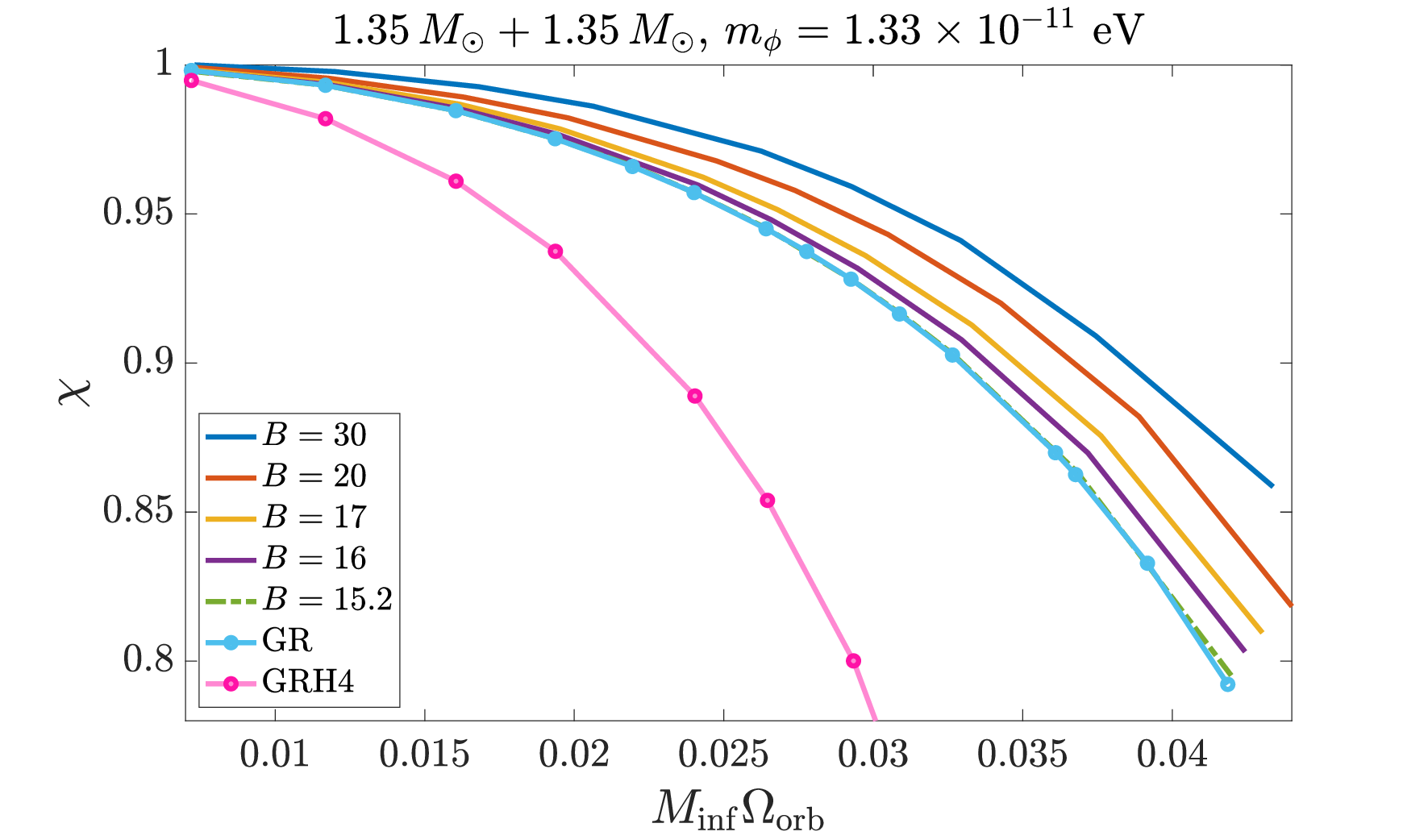}
    \includegraphics[width=\columnwidth]{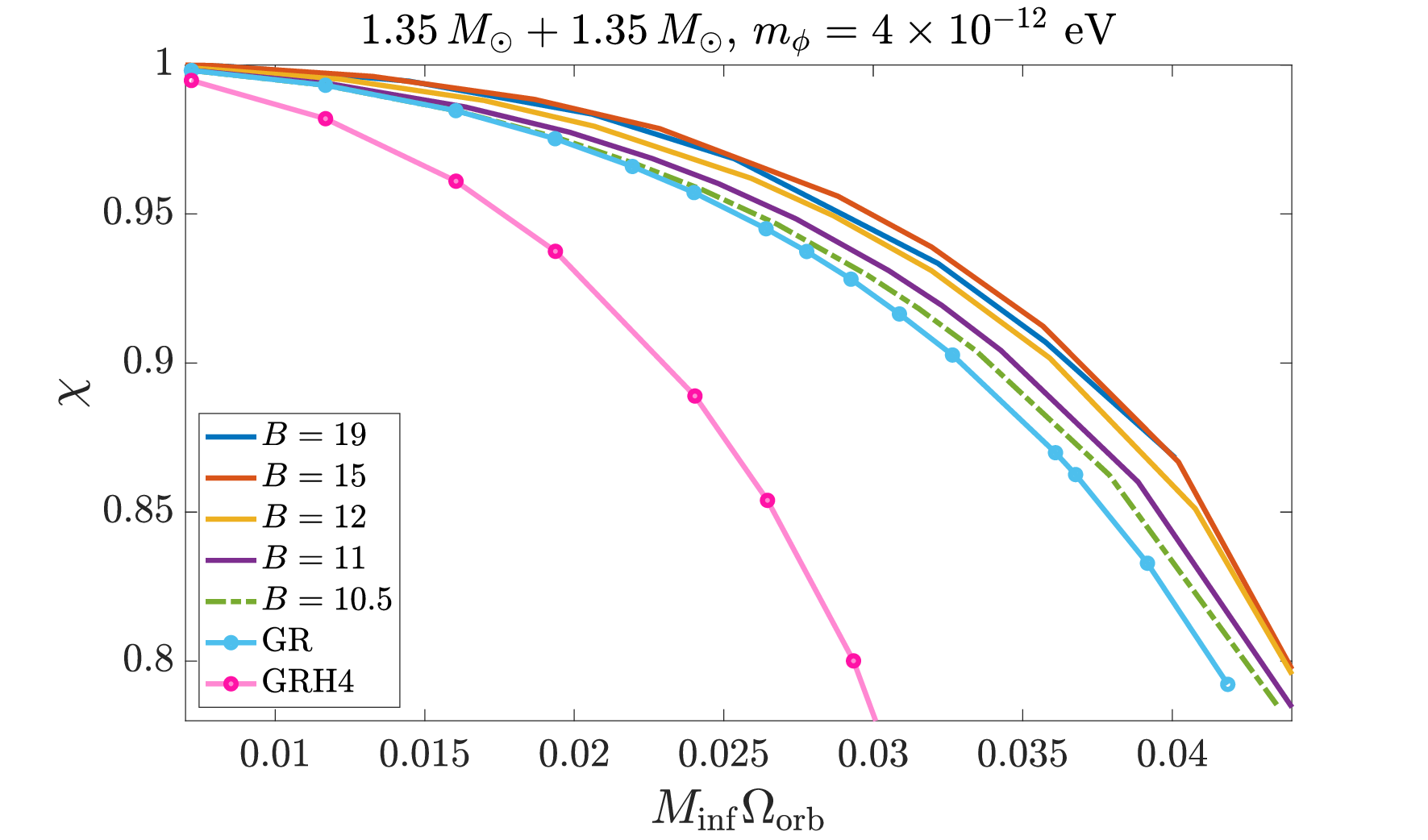}
    \includegraphics[width=\columnwidth]{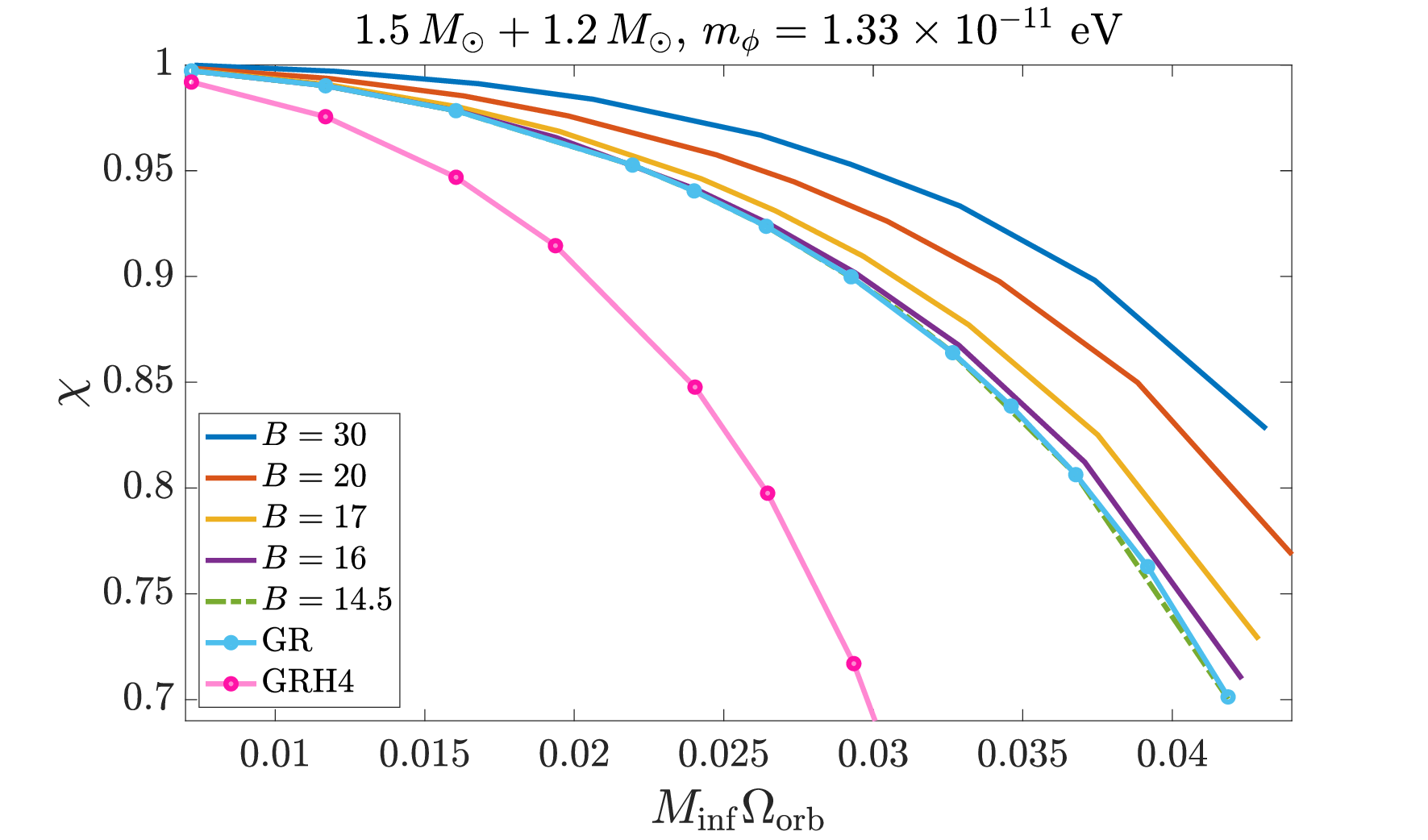}
    \includegraphics[width=\columnwidth]{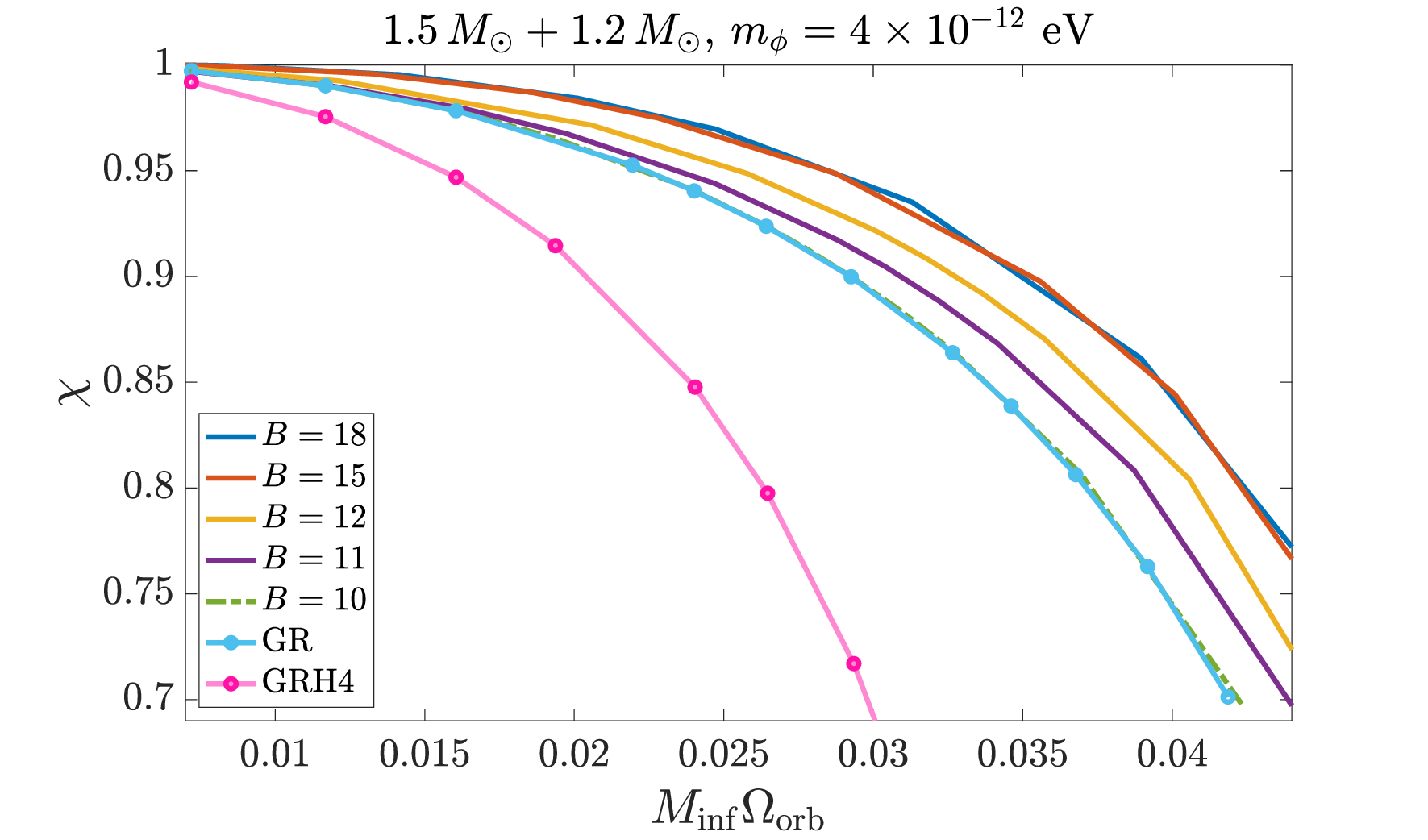}
    \caption{Mass-shedding indicator as a function of orbital frequency for binaries $1.35\,M_\odot+1.35\,M_\odot$ (top) and $1.5\,M_\odot+1.2\,M_\odot$ (bottom). Sequences with dynamical enhancement of the scalarization are shown as solid curves, those with dynamical scalarization as dashed-dotted curves, and the GR sequence is the dotted-solid curve. The ARP4 EOS is employed to model NSs.
    }
    \label{fig:chi}
\end{figure*}

The above conclusion can be further evidenced by looking at the number of cycles, ${\cal N}$, from a given orbital frequency up to merger. Here we estimate ${\cal N}$ in an adiabatic manner by integrating the orbital frequency along the quasi-equilibrium states. Following \cite{tani10,tani14}, we express the energy balance equation as 
\begin{align}
    \frac{dE_{\rm b}}{dt} = -{\cal F},
\end{align}
whereby the orbit shrinks at the rate,
\begin{align}
    \frac{d\Oo}{dt} = -\frac{{\cal F}}{dE_{\rm b}/d\Oo}.
\end{align}
An orbit number of
\begin{align}\label{eq:cycles}
    {\cal N}&=\frac{1}{2\pi} \int \frac{\Oo}{d\Oo/dt}  d\Oo \nn
    &=-\frac{1}{2\pi M_{\rm inf}} \int  \frac{x^{3/2}}{{\cal F}(x)}  \frac{dE_{\rm b}}{dx} dx
\end{align}
will accumulate during the inspiral when the orbital frequency evolves from $\Omega_{\rm i}$ to $\Omega_{\rm e}$, where we introduce $x=(M_{\rm inf}\Oo)^{2/3}$. In numerical integration of Eq.~\eqref{eq:cycles}, we adopt the ansatz (cf.~Eq.~(68) of \cite{tani10}), 
\begin{align}
    E_{\rm b}=E_{\rm b}^{4{\rm PN}}+ax^6+bx^7+cx^8,
\end{align}
to fit the derivative of binding energy with respect to $\Oo$. Here we adopt 4PN result of the binding energy as the principle part ($E_b^{{\rm 4PN}}$; Eq.~(5.6) of \cite{bern18}), and $a$, $b$, and $c$ are the fitting coefficients. In our consideration of energy flux, we ignore the scalar radiation based on two reasons: (i) the emissivity of such radiation is limited over most of the orbital evolution, and (ii) the energy loss via scalar channel is already subdominant to that via traditional GW in the massless ST theory as estimated by \cite{tani14}, let alone the situation in massive ST theory. Therefore, we adopt the 3.5PN approximation for the energy flux, which is given by (Eq.~(314) of \cite{blan14})
\begin{align}
    {\cal F}&=\frac{32}{5}\nu^2x^5\Bigg\{ 1+ \left(-\frac{1247}{336}-\frac{35}{12}\nu\right)x+4\pi x^{3/2} \nn 
    &+\left( -\frac{44711}{9072}+\frac{9271}{504}\nu+\frac{65}{18}\nu^2 \right)x^2 \nn
    &+ \left( -\frac{8191}{672}-\frac{583}{24}\nu \right) \pi x^{5/2} \nn 
    &+\Bigg[ \frac{6643739519}{69854400}+\frac{16}{3}\pi^2-\frac{1712}{105}\gamma_{\rm E} - \frac{856}{105}\ln(16x) \nn 
    &+ \left( -\frac{134543}{7776}+\frac{41}{48}\pi^2 \right)\nu -\frac{94403}{3024}\nu^2 - \frac{775}{324}\nu^3 \Bigg] x^3 \nn 
    &+\left( -\frac{16285}{504} + \frac{214745}{1728}\nu + \frac{193385}{3024}\nu^2 \right)\pi x^{7/2} \Bigg\},
\end{align}
where $\nu=q/(1+q)^2$ is the symmetric mass ratio.

Regulating the upper and lower limits of the integration such that the associated GW frequencies are $f_{\rm gw}\simeq240$~Hz ($M_{\rm inf}\Oo=0.01$) and $f_{\rm gw}\simeq957$~Hz ($M_{\rm inf}\Oo=0.04$) , we list the accumulated GW cycles in Tab.~\ref{tab:cycles}. Consider the almost stiffest and softest EOS that are allowed by GW170817, which are H4 and APR4 \cite{abbo18}, respectively, the number of cycles obtained in GR are 27.45 and 26.24, respectively, for $q=1$, while there are 27.71 and 26.03  cycles for $q=0.8$. Therefore, the uncertainty in the EOS can also be interpreted as the ambiguity of the gravity theory if the resulted ${\cal N}$ in a certain ST theory lies between those for EOS APR4 and H4.

Together with cases with other values of $m_\phi$ not shown in the table, our results are summarized in Fig.~\ref{fig:matome} where the circles and crosses denote the acceptable and unacceptable parameters with respect to the observational results of GW170817. Focusing on systems involving spontaneously scalarized NSs (filled markers), we see that scalar mass of $m_\phi\le10^{-11}$~eV can hardly account for the variation due to EOS, and thus are disfavored after GW170817 in the event that one of NS be spontaneously scalarized.
It is also interesting to note that there are some parameters allowed by $1.5\,M_\odot+1.2\,M_\odot$ binaries are exhibited by equal-mass binaries, and thus more stringent constraint is concluded from the cases with $q=1$. For systems with small mass ratio, the scalarization in the lighter star is much weaker than that in the heavier star, and thus the strength of scalar interaction between the binary is inconsequential. This somehow contradicts the intuitive feeling that one gained from the experience that the more strict constraint is obtained from increasingly asymmetric binaries when analysing the pulsar timing observations in the massless theory, where the emissivity of scalar wave will not be switched off by the scalar mass. Here, instead, the merger is accelerated due to the excess in the lost of orbital energy when developing scalar cloud in the binary.

\subsection{Mass-shedding Criterion}\label{shedding}

The contact of the two NSs could be understood as the moment when one of them loses the feature of being individual. An indicator of such loss of integrity is the formation of a cusp along the direction towards the companion, which can be quantitatively assessed through the ratio between the radial gradient of enthalpy at the pole and at the equatorial point facing the companion \cite{gour01,tani10}. In particular, a dimensionless factor \cite{gour01},
\begin{align}\label{eq:chims}
    \chi_{\rm ms}=\left(\frac{\partial \ln h }{\partial r}\right)_{\rm eq} \left(\frac{\partial \ln h}{\partial r}\right)_{\rm pole}^{-1},
\end{align}
is useful to identify cusp formation: $\chi_{\rm ms}=1$ for static NSs, while $\chi_{\rm ms}=0$ when the cusp is constituted. Since spectral methods cannot resolve well the NS if a cusp is formed at the region closest to the companion, it is unfeasible to construct configuration with $\chi_{\rm ms}\ll1$. In addition, conformal flatness is unlikely to be a fair approximation at very close orbits. In this work, the closest configurations we generated are at a stage less than 1 orbit, i.e., $\lesssim2$~ms, before merger.

Figure \ref{fig:chi} shows the mass-shedding indicator $\chi_{\rm ms}$ as a function of the orbital frequency for the symmetric (top panels) and asymmetric (bottom panels) binaries under our consideration. Several features are observed, including (i) the binaries pertaining to the stiffer EOS H4 start to contact at a lower orbital frequency since the tidal effect is more pronounced; (ii) dynamical scalarization does not affect much $\Oo$ at the onset of mass shedding; (iii) the deformation indicator $\chi$ at a given $\Oo$ is less for increasingly scalarized configuration, which is due to the extra attractive force provided by the scalar field, and is in line with the finding of \cite{shib14} that the central density of NS components keep increasing until merger while a decrease is seen shortly before the merger in GR. However, for the viable ST parameters summarised in Fig.~\ref{fig:matome}, the onset of mass-shedding does not sizeably affected by scalar effects.

\section{Summary and discussion}\label{final}

In order to consistently investigate the constraints that could be obtained from observed gravitational waveforms, detailed understanding of the dynamics during late-inspiral-to-merger is requisite. Owing to the non-linearity manifesting in this regime, numerical-relativity simulation is crucial and serves as the unique tool for this purpose. Constructing quasi-equilibrium states as ID is therefore the first step for the accurate modelling of the gravitational waveforms. We provided reliable ID of binaries consisting of two spontaneously scalarized NSs in massive ST theories since a massless scalar field is excluded by pulsar-timing observations for theories with a high coupling constant $B$.
The scalar mass gives rise to certain hurdles in solving the elliptic-type equation \eqref{eq:ellip_sca} due to the exponentially-decaying behavior of the scalar field [Eq.~\eqref{eq:asymphi}]. An auxiliary scalar field $\xi$ is introduced for better treatment by the spectral code FUKA \cite{pape21}, and is solved for according to the modified equation \eqref{eq:ellip_sca_mod}.

For equilibrium states of binaries generated here, the asymptotic equality \eqref{eq:virial}, dictated by the virial theorem, is met within $0.06$\%, and some of them have been evolved for a few orbits to reaffirm that the quasi-circular motion is guaranteed. The constructed binary configurations thus provide the essential setup for future numerical-relativity study of BNSs in massive ST theories. In addition to future use, qualitative characteristics of the scalar influence can be readily extracted by comparing the equilibria to GR ones. In particular, it is confirmed that the quasi-equilibrium sequences in the ST theory are indistinguishable from that in GR until the orbital separation becomes approximately 3--6 times the Compton length scale of the scalar field, i.e., $a \agt 3$--$6\comp$. Then, at $a \sim 3$--$6\comp$, the enhancement of the scalar field sets in due to the interaction of the scalar clouds of the two NSs (Figs.~\ref{fig:seq} and \ref{fig:seqII}). Accordingly, the gravitational fields will be modified, resulting in the deviation of the quasi-equilibrium sequences from GR. 

To quantify the deviation of sequences in ST from those in GR, we estimate the number of cycles in GWs accumulating over a certain range of orbital frequency [Eq.~\eqref{eq:cycles}]. The tolerance in the stiffness of EOS concluded from GW170817 roughly spans over from the EOS APR4 to H4 \cite{abbo17,abbo19prx}, and thus we adopt the EOS APR4 to derive conservative bounds on the ST parameters, provided that the scalar effects contribute to waveforms in a similar way as tidal effects (Fig.~\ref{fig:cmpr}). We found that the cycles undergone by GWs indeed decrease with a stronger scalar cloud (Tab.~\ref{tab:cycles}) and/or a stiffer EOS. The error budget in ${\cal N}$ defined by the EOS APR4 to H4 can thus be translated to the upper bound on the scalar-induced dephasing in waveforms. Comparing the cycles of ST binaries pertaining to the EOS APR4 to those of GR binaries following the EOS H4, our results are summarised in Fig.~\ref{fig:matome}, where a lower bound of $m_\phi \agt  10^{-11}$\,eV can be reckoned. We also noticed that the most stringent limit is placed by equal-mass binaries, implying that the derived constraint on the scalar mass assuming a spontaneously scalarized NS is in part of the BNS should be robust even though we do not span over a wide range of mass ratio. 
For $m_\phi \agt  10^{-11}$\,eV and a mild coupling strength $B\alt17$, the scalar-cloud interaction effect is not appreciable during the inspiral stage of BNSs despite that both members are scalarized, and can be seen only when the binary is just outside the last stable orbit. The onset of mass-shedding for plausible ST theories essentially matches to the GR cases (cf.~Figs.~\ref{fig:matome} and Fig.~\ref{fig:chi}).

It is important to note yet another layer of complication for the degeneracy between tidal effects, both adiabatic \cite{damo10} and dynamical ones, and the late enhancement of scalarization, either dynamically triggered or through interacting scalar clouds as suggested by Figs.~\ref{fig:seq} and \ref{fig:seqII} (see also \cite{ma23}). It has been known that NSs' tidal response will be modified in ST theories with a massless scalar field so that (i) the tidal effect will appear at 3PN order \cite{damo98} or even at 1PN order \cite{espo09} in the case of dynamical scalarization, (ii) the Love number will increase or decrease depending on the compactness and the ST parameters \cite{pani14,yaza18,brow22}, and (iii) a novel class of Love number is introduced by the scalar field, leading to, e.g., dipolar tidal effects \cite{bern20,geme23}. Relevant studies in the massive ST theory have not been addressed to our knowledge, and a numerical study of scalar-induced modulation in finite-size effects will constitute an essential step toward testing ST theories with GW physics. In this series of investigation, we hope to address this issue to some extent.

\section*{Acknowledgement}
Numerical computation was performed on the clusters Sakura and Cobra at the Max Planck Computing and Data Facility. This work was in part supported by Grant-in-Aid for Scientific Research (grant Nos.~20H00158 and 23H04900) of Japanese MEXT/JSPS.

\bibliographystyle{apsrev4-2}
\bibliography{references}

\end{document}